\newlength{\intwidth}

\documentclass{jfm}
\usepackage{amssymb}

\usepackage{array}
\usepackage{amsmath}
\usepackage{latexsym}
\usepackage{bezier}
\usepackage{psfrag,graphicx}
\usepackage{bm}
\usepackage{float}
\usepackage{multirow}
\usepackage{mathrsfs}
\usepackage{color}

\newcommand{\Natural}{\mathbb{N}}

\begin{document}

\title[High-speed standard magneto-rotational instability]{High-speed standard magneto-rotational instability}
\author{Kengo Deguchi}
\affiliation{
School of Mathematical Sciences, Monash University, VIC 3800, Australia
}

\maketitle

\begin{abstract}
The large Reynolds number asymptotic approximation of the neutral curve of Taylor-Couette flow subject to axial uniform magnetic field is analysed. 
The flow has been extensively studied since early 90's as the magneto-rotational instability (MRI) occurring in the flow could possibly explain the origin of the instability observed in certain astrophysical objects.
Elsewhere the ideal approximation has been used to study high-speed flows, whilst it sometimes produces paradoxical results. 
For example, ideal flows must be completely stabilised for strong enough applied magnetic field, 
but on the other hand the vanishing magnetic Prandtl number limit of the stability should be purely hydrodynamic so instability must occur when Rayleigh's stability condition is violated. 
The first our discovery is that this apparent contradiction can be resolved by showing the abrupt appearance of the hydrodynamic instability at certain critical value of magnetic Prandtl number, which can be found by asymptotically large Reynolds number limit but with long enough wavelength to retain some diffusive effects. 
The second our finding concerns so-called Velikhov-Chandrasekhar paradox, namely the mismatch of the zero external magnetic field limit of the Velikhov-Chandrasekhar stability criterion and Rayleigh's stability criterion. 
We show for fully wide gap cases that the high Reynolds number asymptotic analysis of the MRI naturally yields the simple stability condition that describes smooth transition from Rayleigh to Velikhov-Chandrasekhar stability criteria with increasing Lundquist number.
\end{abstract}

\section{Introduction}

Our concern is with the large Reynolds number fate of the magneto-rotational instability (MRI), which could destabilise hydrodynamically stable rotating flows by an imposed uniform magnetic field in the direction of rotation. 
The possibility of the instability was first pointed out by Velikhov (1959) and Chandrasekhar (1960), who proposed the exceptionally simple stability condition: the azimuthal background flow $V_b(r)$ is stable if
\begin{eqnarray}
\frac{d(r^{-1}V_b)^2}{dr}>0\qquad \text{for all } r,\label{VCcond}
\end{eqnarray}
namely the modulus of the angular velocity of the flow increases outwardly. That stability condition for magnetised flows is weaker than the stability condition for purely hydrodynamic flows by earlier Rayleigh (1917) that the flow is stable if
\begin{eqnarray}
\frac{d(rV_b)^2}{dr}>0\qquad \text{for all } r,\label{Raycond}
\end{eqnarray}
namely the modulus of the angular momentum of the flow increases outwardly. 
Decades later Balbus \& Hawley (1991) spotted that the MRI could destabilise astrophysical flows under Kepler's law, which is stable according to (\ref{Raycond}).
Their discovery surprised other astrophysicists, as turbulence triggered by the instability might explain unknown angular momentum loss in accreting astrophysical objects.


Since Taylor (1923), Taylor-Couette flow has widely been chosen as an ideal testing ground of instabilities appearing in rotating flows because of its simplicity and experimental feasibility. 
The effect of imposed magnetic fields to the stability of the flow has been intensively studied by numerous researchers, as summarised in the recent review by R\"udiger et al. (2018). 
An advantage to use Taylor-Couette flow to study the MRI is that the motion of the independently rotating cylinders can be used to control the laminar flow profile to be within the stable or unstable regime in terms of  (\ref{VCcond}) or (\ref{Raycond}). 
The laminar circular Couette flow is known as one of the simplest Navier-Stokes solutions
\begin{eqnarray}
V_b(r)=R_s r+R_p/r,\label{baseCCF}
\end{eqnarray}
where the constants $R_s$ and $R_p$ are fixed by the cylinder speeds and correspond to the intensity of the solid body rotation part and the potential flow part, respectively. 
The sign of those constants are useful to judge the stability of the flow, since the inequalities appeared in (\ref{VCcond}) and (\ref{Raycond}) become $-V_bR_p>0$ and $V_bR_s>0$, respectively.
The gap between those stability conditions, called anti-cyclonic regime ($R_sR_p>0$), is the most important in astrophysics, at it contains the quasi Kepler rotation.

Both of the stability conditions (\ref{VCcond}) and (\ref{Raycond}) are derived for ideal, axisymmetric perturbations and only represent sufficient conditions of stability. 
Given specific flow configurations, of course more accurate stability condition can be found by numerical computations of linearised viscous resistive magneto-hydrodynamic (MHD) equations.
The earliest computations of magnetised Taylor-Couette flow can be found in Chandrasekhar (1953, 1961), where the inductionless approximation was used to simplify the problem by taking the limit of vanishing magnetic Prandtl number.
Kurzweg (1963) solved finite magnetic Prandtl number problem for the first time, but with rather oversimplified flow geometry and boundary conditions. 
The most complete numerical study at that time was due to Roberts (1964), who proposed realistic boundary conditions and removed the previously used axisymmetric assumption of the perturbations.
The aim of those early computational studies was to explain experimentally observed suppression of the Rayleigh's hydrodynamic instability against the imposed axial uniform magnetic field (Donnelly \& Ozima 1960, 1962; Donnelly \& Caldwell 1964; Brahme 1970).  
The focus of experimental studies then shifted to Rayleigh stable regime after Balbus \& Hawley (1991), but the instability has never been observed for quasi-Kepler rotation. 
A decade later Goodman \& Ji (2002) unveiled the reason why it is so difficult to observe the MRI experimentally, showing  for anti-cyclonic regime that the critical magnetic Reynolds number tends to a constant for small magnetic Prandtl number $P$.
This means that the critical Reynolds number of $O(P^{-1})$ should be enormously large, as the value of $P$ is typically $10^{-5}\sim 10^{-7}$ for the liquid metals used in the experiments.

Nowadays the instability studied by Velikhov (1959) and Chandrasekhar (1960) is called standard MRI because more recently Hollerbach \& R\"udiger (2005) showed that an additional azimuthal component of the external magnetic field dramatically reduces the critical Reynolds number to be $O(P^{0})$.
That new type of instability, called helical MRI, indeed led to the experimental confirmation of the MRI (Stefani et al. 2006, 2007; R\"udiger et al. 2006).

Parallel to those above global studies, so-called local stability analysis has also been used as a conventional tool to explore the parameter space. The key assumption used in the local analysis that the perturbation is locally periodic in all three directions enables us to reduce the stability problem to a single algebraic equation. 
The local analysis is sometimes referred to as the Wentzel-Kramers-Brillouin (WKB) analysis, 
perhaps because the algebraic stability equation derived in the local analysis is equivalent to the eikonal equation governing the WKB phase. 
However, in order to construct a rational global approximation the solutions at each local station must be connected through the WKB amplitude equation at higher order. Furthermore, usually the asymptotic system is not closed unless the WKB solution is matched to the near wall boundary layer solutions. 
In short, the local analysis is actually not mathematically equivalent to the WKB analysis, which serves a rational approximation for global short wavelength modes.
Thus sometimes there is a controversial disagreement between the local and global studies as seen in the helical MRI (Liu et al. 2006; R\"udiger \& Hollerbach 2007; Priede 2011).
Nevertheless, for the standard MRI the local results produce the stability boundary being not too far from the global results (Ji, Goodman \& Kageyama 2001). 
In much the same spirit nonlinear local computations using a periodic shearing box have been intensively performed in astrophysics; see Hawley et al. (1995), Brandenburg et al. (1995) for example.


The aim of this paper is to find accurate self-consistent approximations of the neutral curve assuming large Reynolds number of astrophysical importance. 
The matched asymptotic expansion on which our analysis based is the mathematically rational version of the flow scaling analysis, and has some similarity to the small magnetic Prandtl number analysis of the neutral curve (Chandrasekhar 1953, 1961; Goodman \& Ji 2002; Willis \& Barenghi 2002; Hollerbach \& R\"udiger 2005). 
However, unlike the previous studies here we fix the magnetic Prandtl number $P$ as an $O(1)$ quantity and choose the inverse of the Reynolds number as the intrinsic small parameter of the approximation; in fact $P$ is a constant no matter how small it is, because it should be fixed by the material property of the fluid. 
For purely hydrodynamic Taylor-Couette flow such large Reynolds number analysis successfully produce good approximations of the neutral curve; see Deguchi (2016).


The conclusion made in the many previous asymptotic analyses is, somewhat counterintuitively, the largeness of Reynolds number does not mean that the diffusivity is always negligible. 
As a result, the ideal analyses sometimes produce an outcome contradicting to the large Reynolds number limit of the full equations. 
For example, it is widely known that at least one inflection point of the base flow is necessary to destabilise ideal shear flows, but in reality TollmienÐSchlichting wave type of instability occurs without any inflection point (see Lin 1955, Drazin \& Reid 1981); recently Deguchi (2017) found similar viscous destabilisation in Rayleigh stable cyclonic Taylor-Couette flow.
The most famous paradoxical ideal result in the MRI might be so-called Velikhov-Chandrasekhar paradox, namely the zero external magnetic field limit of the Velikhov-Chandrasekhar condition (\ref{VCcond}) does not coincides with the purely hydrodynamic result (\ref{Raycond}). 
The paradox has been resolved by Kirillov \& Stefani (2011) within the framework of the local analysis, but not yet for the global context. 

In the next section we shall formulate our problem based on viscous resistive MHD equations. 
Section 3 compares the ideal and full numerical results. 
We begin the comparison by the local analysis to show that the ideal result by Velikhov (1959) indeed gives a good approximation for some cases. The results are then extended to narrow and wide gap Taylor-Couette flows.
The stabilisation effect by the ideal analysis competes with the destabilisation effect of the hydrodynamic instability  when the magnetic Prandtl number becomes relatively small, thereby leading the breakdown of the ideal result.
The reason of the breakdown is explained in section 4 using the asymptotic limit with assuming long wavelength of the perturbations. 
Section 5 concerns the asymptotic limit of the MRI, which appears at the Rayleigh stable region. 
The simple stability criterion to be derived in that section indeed describes the smooth transition from Rayleigh's condition (\ref{Raycond}) to the Velikhov-Chandrasekhar condition (\ref{VCcond}) with increasing the magnetic field from zero. 
Finally, in section 6, we conclude with a brief discussion.

\section{Formulation of the problem}
Consider non-dimensional incompressible viscous resistive MHD equations
\begin{subequations}\label{MHDeq}
\begin{eqnarray}
(\partial_t+\nabla \cdot \mathbf{v})\mathbf{v}-(\mathbf{b}\cdot \nabla )\mathbf{b}=-\nabla q+\nabla^2 \mathbf{v},\label{MHDmo}\\
(\partial_t+\nabla \cdot \mathbf{v})\mathbf{b}-(\mathbf{b}\cdot \nabla )\mathbf{v}=P^{-1}\nabla^2 \mathbf{b},\\
\nabla \cdot \mathbf{v}=0,\qquad \nabla \cdot \mathbf{b}=0
\end{eqnarray}
\end{subequations}
in the cylindrical coordinates $(r,\theta,z)$. Here we write the velocity vector as $\mathbf{v}=u\mathbf{e}_r+v\mathbf{e}_{\theta}+w\mathbf{e}_z$, the magnetic vector as $\mathbf{b}=a\mathbf{e}_r+b\mathbf{e}_{\theta}+c\mathbf{e}_z$ and the total pressure as $q$.
The magnetic Prandtl number $P$ is the ratio of the kinematic viscosity to the magnetic diffusivity. 
Here we scale the magnetic field so that the pre-factor of the Lorentz force term $(\mathbf{b}\cdot \nabla) \mathbf{b}$ in (\ref{MHDmo}) is normalised. 
Also the normalised viscous term $\nabla^2 \mathbf{v}$ in the same equation means that our choice of the velocity scale is the viscous one, and hence the Reynolds numbers should appear in the base flow.
The flow is assumed to be enclosed by infinitely long co-axial cylinders placed at $r=r_i, r_o$, where $r_o>r_i$. 
Our length scale is the half gap of the cylinders, namely $r_o-r_i=2$.
Thus using the radius ratio $\eta=r_i/r_o$, we can express the positions of the cylinders as
\begin{eqnarray}
r_i=\frac{2\eta}{1-\eta},\qquad r_o=\frac{2}{1-\eta}.
\end{eqnarray}
Throughout the paper we assume the axisymmetry of the flow.

Rotating the cylinder walls with constant angular speeds, well-known Taylor-Couette flow is realised. 
We further apply uniform axial magnetic field of non-dimensional magnitude $B_0$ to trigger the MRI. 
Note that we have to scale it by the magnetic Prandtl number to get the Hartmann and Lundquist numbers
\begin{eqnarray}
H=B_0P^{1/2}, \qquad S=B_0P,
\end{eqnarray}
respectively.
The base flow can be written as
\begin{eqnarray}
\mathbf{v}=V_b(r)\mathbf{e}_{\theta}, \qquad \mathbf{b}=B_0 \mathbf{e}_z,
\end{eqnarray}
where the laminar velocity field of Taylor-Couette flow $V_b$ is sum of the solid body rotation part and the potential flow part as shown in (\ref{baseCCF}).  
In order to satisfy
\begin{eqnarray}
V_b(r_i)=R_i,\qquad  V_b(r_o)=R_o,
\end{eqnarray}
where $R_i$ and $R_o$ are the Reynolds numbers associated with the inner and outer cylinder speeds, respectively, we must set
\begin{eqnarray}
R_s=\frac{R_o-\eta R_i}{2(1+\eta)},\qquad R_p=\frac{\eta^{-1}R_i- R_o}{2(1+\eta)}r_i^2.\label{defRsRp}
\end{eqnarray}
Kepler's law states that square of period of the orbital motion should proportional to $r^3$. 
If we apply this condition for all streamlines then we have the base flow $V_b(r)\propto r^{-1/2}$, which is never realisable as a base flow.
Thus in Taylor-Couette flow studies Kepler's law is only applied for the cylinders' motion. 
The base flow in this case, satisfying
\begin{eqnarray}
R_i/R_o=\eta^{-1/2}, \label{Kepler}
\end{eqnarray}
is called the quasi-Keplerian rotation.

The linear perturbation to the base flow can be expressed by normal modes so we write
\begin{eqnarray}
\mathbf{v}=V_b(r)\mathbf{e}_{\theta}+\widehat{\mathbf{v}}(r)\exp(\sigma t+ikz), \qquad \mathbf{b}=B_0 \mathbf{e}_z++\widehat{\mathbf{b}}(r)\exp(\sigma t+ikz),\label{linearexp}
\end{eqnarray}
where $k$, $\sigma$ are the axial wavenumber and the complex growth rate of the perturbation, respectively.
Substituting (\ref{linearexp}) to (\ref{MHDeq}) and linearising the equations, 
\begin{subequations}\label{eqcomp}
\begin{eqnarray}
\left[ \begin{array}{c}  -2r^{-1}V_b v\\ r^{-1}(rV_b)'u \\ 0 \end{array} \right]
=-\left[ \begin{array}{c}  Dq\\ 0 \\ikq \end{array} \right]
+  \left[ \begin{array}{c}  (DD_*-k^2-\sigma)u\\ (DD_*-k^2-\sigma)v\\ (D_*D-k^2-\sigma)w \end{array} \right]
+ikB_0
\left[ \begin{array}{c}  a\\ b\\ c \end{array} \right], 
\\
\left[ \begin{array}{c} 0\\ -r(r^{-1}V_b)'a\\ 0 \end{array} \right]=
P^{-1} \left[ \begin{array}{c} (DD_*-k^2-\sigma)a\\ (DD_*-k^2-\sigma)b\\  (D_*D-k^2-\sigma)c \end{array} \right]
+ikB_0
\left[ \begin{array}{c} u\\ v\\ w  \end{array} \right],\label{compind}
 \\
D_*u+ikw=0,\qquad D_*a+ikc=0.~~~~
\end{eqnarray}
\end{subequations}
Here we have removed the hats for the sake of simplicity and $D=\partial_r,D_*=\partial_r+1/r$ are the usual Chandrasekhar notations. 
Dropping all the diffusive terms form (\ref{eqcomp}) and assuming $u$ vanishes on the walls, Chandrasekhar (1960) derived an integral equation
\begin{eqnarray}
(\sigma^2+k^2B_0^2)^2 \int^{r_o}_{r_i} r(|D_*u|^2+k^2|u|^2) dr~~~~~~~~~~~~~~~~~~~\nonumber \\
~~~~~~~~~+k^2\int_{r_i}^{r_o} \left (\sigma^2\frac{1}{r^2}\frac{d(rV_b)^2}{dr}+k^2B_0^2r^2\frac{d(r^{-1}V_b)^2}{dr} \right ) |u|^2 dr=0,
\end{eqnarray}
where $\sigma^2$ should be real. 
The Velikhov-Chandrasekhar result follows by noting that if (\ref{VCcond}) holds then $\sigma^2<0$ and hence there is no instability for the ideal flow.
(We also remark that if (\ref{Raycond}) holds and $B_0=0$ then $\sigma^2<0$.)


Now let us introduce the potentials satisfying $u=ik\phi$, $w=-(r\phi)'/r$, $a=ik\psi$, $c=-(r\psi)'/r$ to convert (\ref{eqcomp}) to the simpler form
\begin{subequations}\label{potwide}
\begin{eqnarray}
(\sigma-\triangle)\triangle \phi -ik B_0 \triangle \psi-ik(2R_s+2r^{-2}R_p) v=0,\\
(\sigma-\triangle)v-ik B_0 b+ik2R_s\phi=0,\\
(\sigma-P^{-1}\triangle)\psi-ik B_0 \phi=0,\\
(\sigma-P^{-1}\triangle)b-ik B_0 v+ik2r^{-2}R_p\psi=0,
\end{eqnarray}
\end{subequations}
where $\triangle=DD_*-k^2$. 
The no-slip conditions on the walls are satisfied if
\begin{eqnarray}
\phi=D\phi=v=0\qquad \text{at} \qquad r=r_i, r_o.\label{noslip}
\end{eqnarray}
For the magnetic field, perfectly conducting or insulating walls are frequently considered; see R\"udiger et al. (2018) and references therein. If we assume that the walls are made of perfectly conducting materials the boundary conditions are 
\begin{eqnarray}
\psi=D_*b=0\qquad \text{at} \qquad r=r_i, r_o,\label{conducting}
\end{eqnarray}
whilst for the perfectly insulating walls we must impose the conditions
\begin{subequations}\label{insulating}
\begin{eqnarray}
\frac{I_1(kr_i)}{I_0(kr_i)}D_*\psi-k\psi=b=0\qquad \text{at} \qquad r=r_i,\\
\frac{K_1(kr_o)}{K_0(kr_o)}D_*\psi+k\psi=b=0\qquad \text{at} \qquad r=r_o.
\end{eqnarray}
\end{subequations}
Here $I_0,I_1$ are the zeroth and first modified Bessel functions of the first kind, $K_0,K_1$ are the zeroth and first modified Bessel functions of the second kind. 

Before we begin the large Reynolds number asymptotic analysis we note some known asymptotic properties of the stability in the limit of $P\rightarrow 0$. 
Chandrasekhar (1953; 1961) took this limit by assuming $V_b, H\sim O(P^0)$, $\mathbf{u}\sim O(P^{0})$, $\mathbf{b}\sim O(P^{1/2})$ and showed that the term at the left hand side of (\ref{compind}) should drop from the leading order system. 
Goodman \& Ji (2002) proved that without that term axisymmetric instability in the anticyclonic regime is not possible for the narrow-gap cases, and subsequently the proof was extended for the wide-gap cases by Herron \& Goodman (2006). 
Note that the helical MRI could be stimulated at the limit due to some other extra terms involving the azimuthal background magnetic field (Hollerbach \& R\"udiger 2005). 

Goodman \& Ji (2002) found another possible small $P$ limit where the important term mentioned above plays a role.
In order to find the reduced system at the limit, it is convenient to apply the scaling
\begin{eqnarray}
V=B_0 R_p^{-1}v,\qquad \Phi=PB_0 \phi,\qquad B=P^{-1}R_p^{-1}b,\qquad C=\frac{R_pP^2B_0^2}{R_s}\label{scaledV}
\end{eqnarray}
to convert the neutral and steady version of (\ref{potwide}) to
\begin{subequations}\label{scaled}
\begin{eqnarray}
\frac{\triangle^2 \Phi}{B_0^2R_p^2P^3} +ik \frac{\triangle \psi}{R_p^2P^2}+ik\left (\frac{1}{C}+\frac{r_o^2}{r^2B_0^2P^2}\right ) \frac{2V}{r_o^2}=0,\label{moscaled1}\\
\frac{\triangle V}{B_0^2P}+ik B-ik\frac{2\Phi}{Cr_o^2}=0,\label{moscaled2}\\
\triangle \psi+ik \Phi=0,\label{indscaled1}\\
\triangle B+ik V-ik\frac{2\psi}{r^2}=0.\label{indscaled2}
\end{eqnarray}
\end{subequations}
Here we assume that the scaled quantities (\ref{scaledV}) and $k$ are $O(P^0)$.
The purpose of the scaling is to normalise the induction equations (\ref{indscaled1}), (\ref{indscaled2}). 
The last term in the left hand side of (\ref{indscaled2}), corresponding to the important term spotted by Goodman, Ji and Herron, never drops under this rescaled system. 
Goodman \& Ji (2002) kept the magnetic Reynolds number $R_m=PR_p$, the Lundquist number and $R_s/R_p$ as $O(P^0)$ quantities to show that only the viscous terms in (\ref{moscaled1}) and (\ref{moscaled2}) drop from the leading order equations. 
As those terms possess the highest derivative, their inductionless limit corresponds to a singular asymptotic limit where the hydrodynamic boundary conditions cannot be satisfied. 
As a result passive near wall boundary layers matching to the inviscid outer solution should appear as shown by Goodman \& Ji (2002). 

When the base flow is close to the pure potential flow, Willis \& Barenghi (2002) showed that the stability is scaled in the Hartmann number rather than the Lundquist number. The corresponding asymptotic limit can be taken by assuming  $H\sim O(P^0)$. 
In this case the only small term drops out of the equations is the term proportional to $1/C$ in (\ref{moscaled1}), and hence this is a regular asymptotic limit. From the assumptions we can find that $R_s/R_p$ must be $O(P)$, namely the flow is close to the potential flow (i.e. close to the Rayleigh line which separates the unstable/stable regions in the sense of Rayleigh).

\section{The full versus ideal stability analyses}

As appreciated by many hydrodynamic researchers, the largeness of the Reynolds number does not immediately mean that we can neglect the diffusivities. Here we shall clarify when do the ideal results give a good approximation of the full viscous resistive MHD results.

\subsection{The narrow-gap limit}
In the majority of this section we explore the parameter space taking the narrow-gap limit $\eta \rightarrow 1$.
The result of that simplified problem shall then be extended to the wide gap cases later. 
Let us write the gap coordinate as $x=r-r_m$ with the reference radius $r_m\in [r_i,r_o]$. The choice of $r_m$ is rather arbitrary but in our numerical computation it is fixed to be the mid gap $r_m=(r_i+r_o)/2$ for the sake of definiteness.
While taking the limit $(1-\eta) \rightarrow 0$, we keep the shear Reynolds number
\begin{subequations}\label{ngparam}
\begin{eqnarray}
R=-r(r^{-1}V_b)'|_{r=r_m}=\frac{2R_p}{r_m^2},
\end{eqnarray}
and the inverse Rossby number
\begin{eqnarray}
\omega=\frac{2(r^{-1}V_b)|_{r=r_m}}{R}=\frac{2R_s}{R}+1,\label{Rossby}
\end{eqnarray}
\end{subequations}
as $O((\eta-1)^0)$ quantities.
The other variables/parameters $x,\sigma,k,B_0,P$ are also assumed to be $O((\eta-1)^0)$ quantities but $r_m\sim O(1/(1-\eta))$ is very large. 

The limiting form of (\ref{potwide}) as $\eta \rightarrow 1$ can be found as
\begin{subequations}\label{potnarrow}
\begin{eqnarray}
(\sigma-\triangle)\triangle \phi -ik B_0\triangle \psi-ikR\omega v=0,\label{ng1}\\
(\sigma-\triangle)v-ik B_0b+ikR(\omega-1)\phi=0,\label{ng2}\\
(\sigma-P^{-1}\triangle)\psi-ik B_0\phi=0,\label{ng3}\\
(\sigma-P^{-1}\triangle)b-ik B_0 v+ikR\psi=0,\label{ng4}
\end{eqnarray}
\end{subequations}
where $\triangle=\partial_x^2-k^2$. 
From (\ref{noslip}), (\ref{conducting}), (\ref{insulating}) the boundary conditions at the limit are
\begin{eqnarray}
\phi=\phi'=v=\psi=b'=0\qquad \text{at}\qquad x=\pm 1 \label{ngcd}
\end{eqnarray}
for the perfectly conducting walls, and
\begin{eqnarray}
\phi=\phi'=v=\psi'\pm k\psi=b=0\qquad \text{at}\qquad x=\pm 1 \label{ngin}
\end{eqnarray}
for the perfectly insulating walls. Here the prime denotes differentiation with respect to $x$.



\begin{figure}
\centering
\includegraphics[scale=1.2]{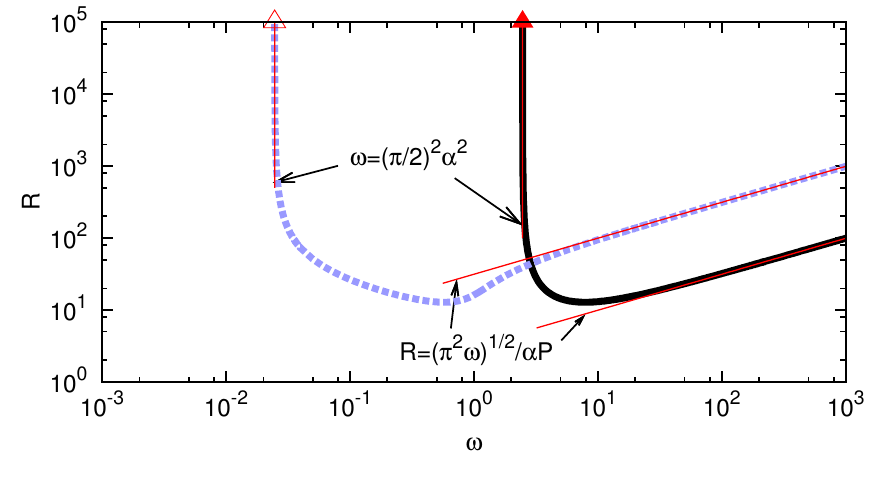} \\
\includegraphics[scale=1.2]{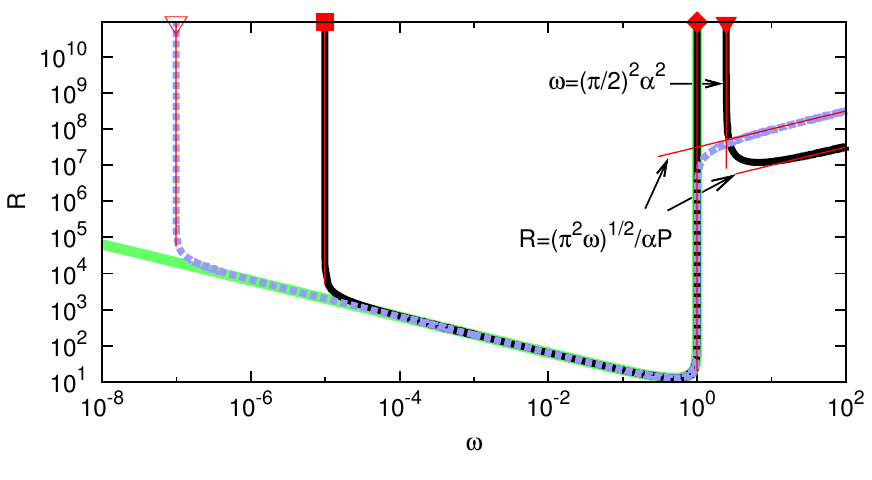} 
\caption{The neutral curves for the shearing box computation. The black solid and blue dashed curves are the results for $\alpha=1$ and $\alpha=0.1$, respectively. Top panel: $P=1$, bottom panel: $P=10^{-6}$. 
The thin red lines are the asymptotic results; see (\ref{ngidealpd}) for the ideal limit, (\ref{hosb}) for the high-rotation limit and figure 5 for the long wavelength limit. 
The thick green curve in the bottom panel is the purely hydrodynamic result (\ref{ngidealpdh}).
}
\label{fig:longwave}
\end{figure}


If we instead use the periodic boundary conditions in $y$ then the narrow-gap problem becomes identical to the local analysis or equivalently the two-dimensional version of the shearing box computation.
As remarked earlier we have to keep in mind that it is not of mathematically consistent approximation of the global solution because there are two important ingredients missing, namely the effects of curvature and the boundary conditions. 
Nevertheless, the local analysis sometimes has ability to capture some qualitative feature of global stability results.
In the next two subsections we shall compare the shearing box computation and the global analysis of the narrow-gap problem. 
In order to make a better comparison to the other boundary conditions we set the periodic boundaries at $y=\pm 2$. Then steady shearing box solutions may also satisfy
\begin{eqnarray}
\phi=\phi''=v=\psi=b=0\qquad \text{at}\qquad x=\pm 1\label{ngpd}
\end{eqnarray}
which is similar to (\ref{ngcd}); the conditions for the magnetic field is unchanged from the perfectly conducting case, whilst the velocity parts are required to satisfy free stress conditions on the walls rather than the no-slip conditions.

\subsection{The shearing box computations}


Velikhov (1958) found for ideal flows that too strong magnetic field applied should stabilise the MRI. 
In order to derive the ideal result first we neglect all the diffusive effects from equations (\ref{potnarrow}). The neutral ideal perturbations are governed by the single combined equation
\begin{eqnarray}
\triangle \psi +\frac{\omega}{\alpha^2} \psi=0, \label{diffidealng}
\end{eqnarray}
where
\begin{eqnarray}
\alpha=\frac{B_0}{R}
\end{eqnarray}
is the inverse magnetic Mach number.
The solution is simply a sinusoidal function
\begin{eqnarray}
\psi=\cos\left (\frac{n\pi}{2} x\right ),\qquad \frac{\omega}{\alpha^2}-k^2=\frac{n^2\pi^2}{4},\qquad n\in \Natural.\label{idealBH}
\end{eqnarray}

In order to observe the ideal stability boundary in the full stability calculations it is convenient to fix $\alpha$ rather than $B_0$. 
The use of the sinusoidal function in $y$ convert (\ref{potnarrow}) into
the single algebraic equation
\begin{eqnarray}
(P^{-1}l^4+(kR\alpha)^2)^2l^2+(kR P^{-1})^2\omega (\omega-1)l^4-\omega \alpha^2(kR)^4=0,\label{comba}
\end{eqnarray}
where $l=\sqrt{k^2+(n\pi/2)^2}$, $n\in \Natural$.
Here we have assumed $\sigma=0$ because Ji et al. (2001) proved that the neutral axisymmetric neutral modes are always steady. 

It is important to note here that the purely hydrodynamic result 
\begin{eqnarray}
l^6+k^2R^2\omega (\omega-1)=0\label{combh}
\end{eqnarray}
can be recovered if we set $\alpha=0$ or $P=0$ in (\ref{comba}).  
The hydrodynamic neutral mode is only possible for $\omega \in [0,1]$.


The algebraic equation (\ref{comba}) has been studied by many researchers to find simple conventional stability conditions (see Sano \& Miyama 1999; Ji, Goodman \& Kageyama 2001; Velikhov et al. 2006; Kirillov \& Stefani 2011, for example). The key observation that made the previous local analyses so simple is that the shearing box problem does not have particular spatial length scale. In fact the equation (\ref{comba}) is invariant under the transformation $(R,\alpha,k,n)\rightarrow (a^2R,a^{-1}\alpha,ak,an)$ for any constant $a$, which is the degree of freedom due to the free length scale. As a result, the stability conditions so far use the parameters normalised by the wavenumber. 
However, those stability conditions are not applicable for our neutral curve because there is the maximum flow scale in the $y$ direction. 
In that case we can no more use the normalised parameters and, more importantly, the wavenumber dependence of the stability curve must be eliminated through the optimisation process to seek the most dangerous mode.

The easiest way to find such optimised neutral curve for given $P$, $\alpha$, $\omega$ is to solve equation (\ref{comba}) for $R^2$ which must be minimised against the wavenumbers $k,n$. 
The neutral curves of the shearing box computations are summarised in figure 1. 
The stability diagram is point symmetric in $\omega$--$R$ plane so hereafter we only consider positive $R$. The flow is always stable when $\omega$ is negative so this region is not shown throughout the paper. 
At the narrow gap limit the inequalities (\ref{VCcond}) and (\ref{Raycond}) become $\omega<0$ and $\omega(1-\omega)<0$, respectively.
There are two Rayleigh stable regions, cyclonic ($\omega<0$) and anticyclonic ($\omega>1$) regimes, where the Kepler rotation $\omega=4/3$ (see (\ref{Kepler}), (\ref{Rossby})) belongs to the latter.

For large to moderate magnetic Prandtl numbers the shape of the neutral curves is rather simple as can be seen in the top panel where $P=1$ is chosen. There are two curves shown for $\alpha=1$ (black solid curve) and $\alpha=0.1$ (blue dashed curve). 
Above the neutral curves, the corresponding flow becomes linearly unstable. 
Decreasing $\alpha$ the left hand branch is destabilised, while the right hand branch is stabilised. 
Both branches seem to converge towards some large Reynolds number asymptotic states, which are our main interest here. 
The behaviour of the left hand branch can be explained by the ideal result (the right hand branch converges to the high-rotation limit to be derived in section 5).
The optimised ideal mode at $(k, n)=(0,1)$ found by (\ref{idealBH}) gives the stability boundary
\begin{eqnarray}
\frac{\omega}{\alpha^2}=\left (\frac{\pi}{2}\right )^2,\label{ngidealpd}
\end{eqnarray}
which corresponds to the thin red vertical lines in the figure. 
The larger the normalised magnetic field $\alpha$ is, the larger the critical value of $\omega$ should be.
Thus (\ref{ngidealpd}) indeed describes the Velikhov's stabilisation mechanism by the external magnetic field. 

The ideal result (\ref{ngidealpd}) is independent of $P$ so one may expect to see the similar agreement for small values of $P$ of experimental interest. 
However, we have remarked earlier that at the vanishing magnetic Prandtl number limit the stability of the flow should become purely hydrodynamic. 
At first glance those two statements might look inconsistent, because on one hand the ideal limit predicts that there is no instability below the critical value of $\omega$, which is large when the strong magnetic field is applied, but on the other hand hydrodynamic instability should occur for $\omega \in [0,1]$. 


The behaviour of the neutral curve for smaller magnetic Prandtl number can be found in the bottom panel of figure 1. 
Here we choose $P=10^{-6}$ to compute the stability. 
The solid curve for $\alpha=1$ shows the typical neutral curve when the magnetic field is strong.
There are actually two islands of instability, the right of which corresponds to the similar result as the previous panel and indeed the ideal result predicts the behaviour of its left hand branch. 
When the applied magnetic field is weaker and the ideal limit value of $\omega$ is smaller than unity the ideal limit is not observable; see the dashed curve for $\alpha=0.1$ in the lower panel of figure 1, where the two islands of the instability merge.
The neutral curve appeared at the Rayleigh unstable region might be explained by the hydrodynamic instability (\ref{combh}). In fact, for not too large $R$ the curve can be approximated by the hydrodynamic result plotted by the thick green curve in the figure. This curve can be found by optimising (\ref{combh}) as
\begin{eqnarray}
R^2\omega (1-\omega)=T,\label{ngidealpdh}
\end{eqnarray}
where $T$ is the well-known critical Taylor number $27\pi^4/64\approx 41.094$. The value of $T$ and some key numbers to be derived in the asymptotic analysis are summarised in table 1.
\begin{table}
  \begin{center}
    \begin{tabular}{ccccc}
     Boundary conditions & $T$ &$\omega/\alpha^2$ & $S^2/\omega$ & $R_m$  \\
     Shearing box &  41.094 &  2.467 & 9.870 & 12.821  \\
     Perfectly conducting &106.73& 2.467 & 3 & 6  \\
     Perfectly insulating &106.73& 0 & 6.598 & 3.774 \\
    \end{tabular}
  \end{center}
\caption{The key numbers in the asymptotic analysis for the narrow-gap cases. 
$T$: the critical Taylor number for the hydrodynamic limit (\ref{ngidealpdh}). 
$\omega/\alpha^2$: the critical value for the ideal limit (see (\ref{ngidealpd}) and section 3.3).
$S^2/\omega$: the critical value for the high-rotation limit (see (\ref{hosb}), (\ref{cdholine}) and (\ref{inholine})).
$R_m$: the critical magnetic Reynolds number for the high-rotation limit for all possible external magnetic fields (see (\ref{Rmpd}) and Appendix B). } 
\label{sample-table}
\end{table}

However, it is not clear how the neutral curve in the top panel deforms into that in the bottom panel at this stage.
For large enough Reynolds numbers the hydrodynamic approximation of the instability fails and the neutral curve tends to some constant $\omega$. 
We shall show in section 4 that the asymptotic analysis of this unknown large Reynolds number states yields the critical value of $P$ at which the instability at the Rayleigh unstable region abruptly appears. 
That new asymptotic limit occurs when we choose small wavenumber of $O(R^{-1})$. 
We have seen that the optimum of the ideal limit also occurs at small $k$, but we will see that it should still be larger than $O(R^{-1})$ in asymptotic sense.

\subsection{The perfectly conducting and insulating boundary conditions}

\begin{figure}
\centering
\includegraphics[scale=1.2]{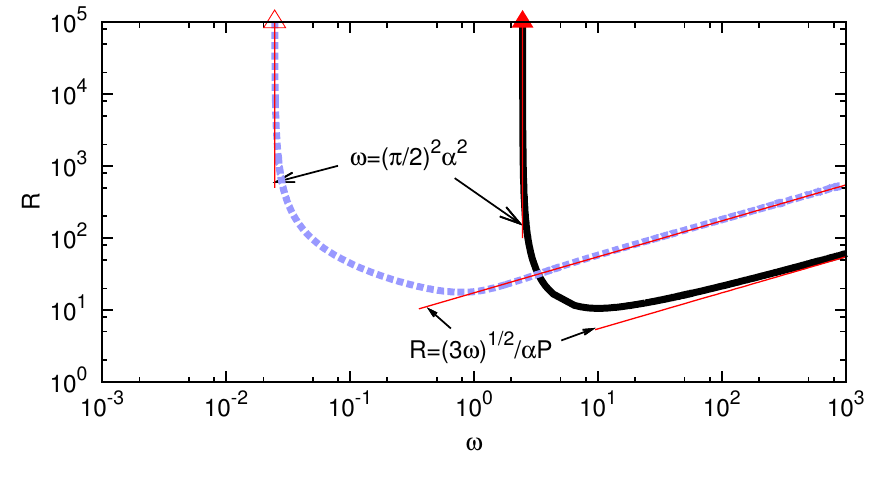} \\
\includegraphics[scale=1.2]{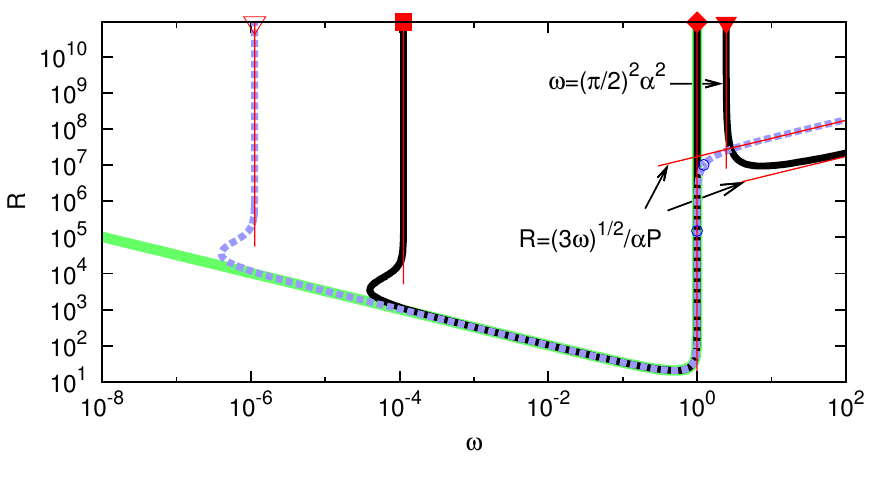} 
\caption{The same picture as figure 1 but for the perfectly conducting boundary conditions at the narrow-gap limit. The thin red lines are the asymptotic results; see (\ref{ngidealpd}) for the ideal limit, (\ref{cdholine}) for the high-rotation limit and figure 7 for the long wavelength limit. The blue circles correspond to the onset of the oscillatory mode.}
\label{fig:longwave}
\end{figure}

\begin{figure}
\centering
\includegraphics[scale=1.2]{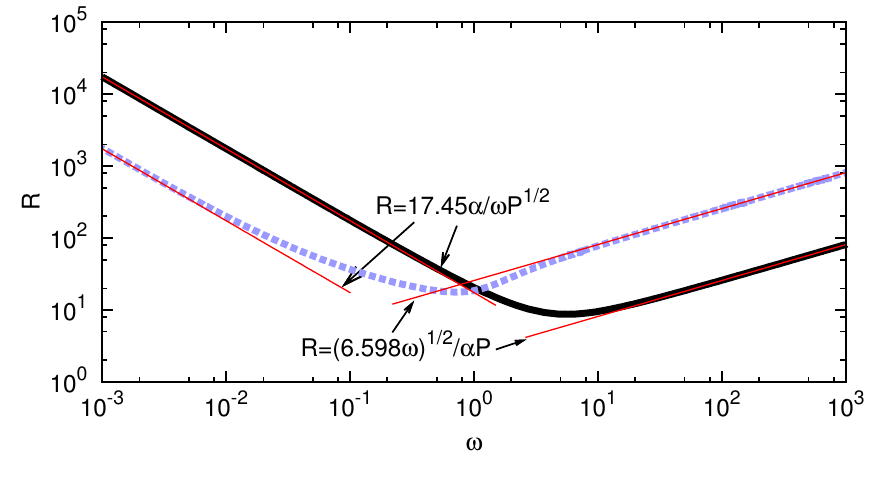} \\
\includegraphics[scale=1.2]{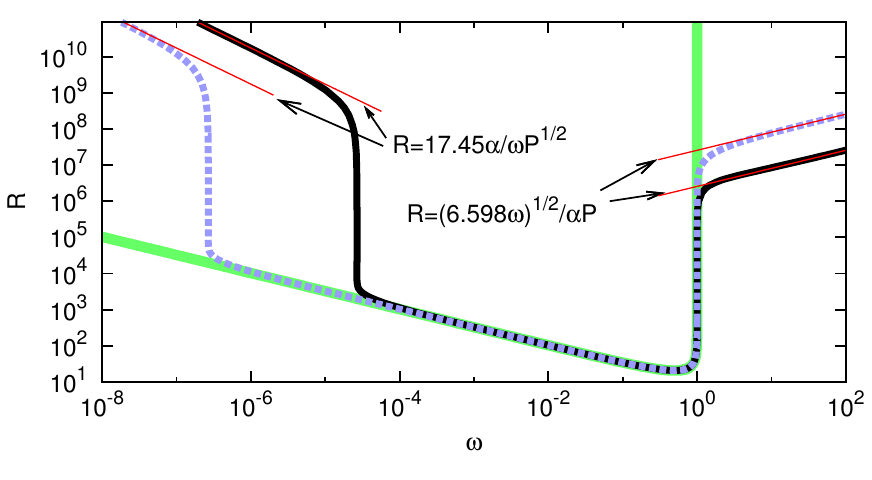} 
\caption{The same picture as figure 1 but for the perfectly insulating boundary conditions at the narrow-gap limit.  The thin red lines are the asymptotic results; see (\ref{inlong}) for the long wavelength limit, (\ref{inholine}) for the high-rotation limit.}
\label{fig:longwave}
\end{figure}
For the other two boundary conditions (\ref{ngcd}) and (\ref{ngin}) we must numerically solve differential equations (\ref{potnarrow}). The equations are first discretised by substituting the modified Chebyshev expansions and evaluating them at the collocation points; see Appendix A. 
Then the resultant linear eigenvalue problem for the eigenvalue $\sigma$ can be solved by LAPACK routines.
As the previous section we fix $\alpha$, $P$ as some constants and seek the neutral curve in the $\omega$-$R$ plane. The largest real part of the eigenvalues must be maximised against $k$ to find the most dangerous mode. The zero locus of the optimised value can then be found by bisection to draw the neutral curve. The numerical resolutions are tested using up to 600 Chebyshev modes.

Figure 2 shows the neutral curve for the perfectly conducting walls. 
The first thing to note is that this figure is overall similar to figure 1, meaning that the `local approximation' captures some qualitative aspects of the global stability result when the cylinders are perfectly conducting and the gap between them is narrow.  
In particular, the identical ideal limits can be seen, as shown in the upper panels.
This is because the ideal solution (\ref{idealBH}) for the most dangerous mode $n=1$ also satisfies the perfectly conducting boundary conditions for $\psi$ shown in (\ref{ngcd}). 
(The other boundary conditions can be accounted by inserting thin passive near wall boundary layers similar to section 5.)
The results shown in the lower panel for smaller $P$ again indicate the emergence of the instability at the Rayleigh unstable region. For not too large Reynolds numbers it can be approximated by the hydrodynamic result (\ref{ngidealpdh}) but with $T\approx 106.73$ for no-slip boundaries (Taylor 1923).


Figure 3 is the results for the perfectly insulating conditions where we can also find the similar convergence to the hydrodynamic result for small $P$ limit. 
However, for large $R$, the asymptotic property of the neutral curve is qualitatively different from the previous two cases. The most striking feature of the upper panel is the absence of the ideal limit, which served accurate threshold value of $\omega$ below which the flow is stabilised. 
The threshold value for the perfectly insulating case is actually predicted to be zero, because the optimised solution of the reduced differential equation (\ref{diffidealng}) with the conditions for $\psi$ shown in (\ref{ngin}) is merely $\omega=k=0, n=1, \psi=1$ for any $\alpha$. 
Hence the neutral curve should tend to $\omega=0$ for large Reynolds numbers, but before reaching this limit the left hand branches of the neutral curves show unknown asymptotic trend where $R\omega$ seems to converge some constant. 
The detailed asymptotic analysis of this state will be studied in section 4. 


\subsection{The wide gap cases}
\begin{figure}
\centering
\includegraphics[scale=1.2]{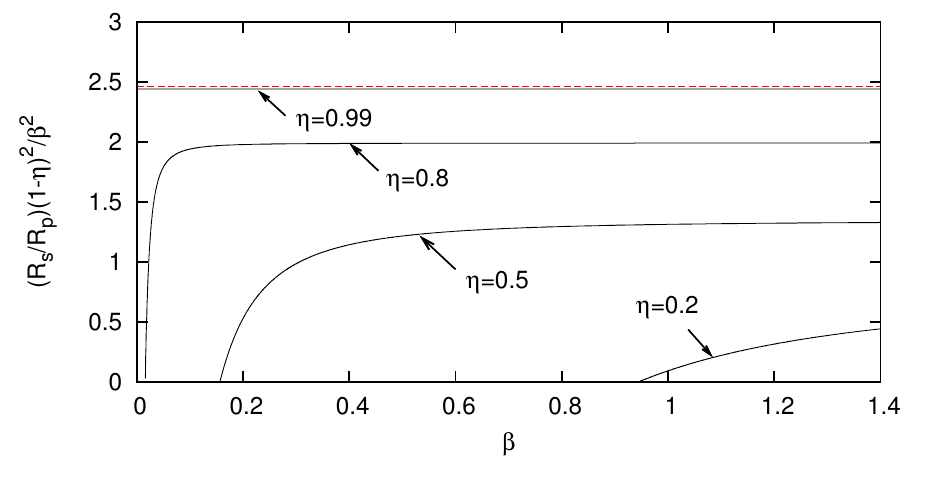} \\
\includegraphics[scale=1.2]{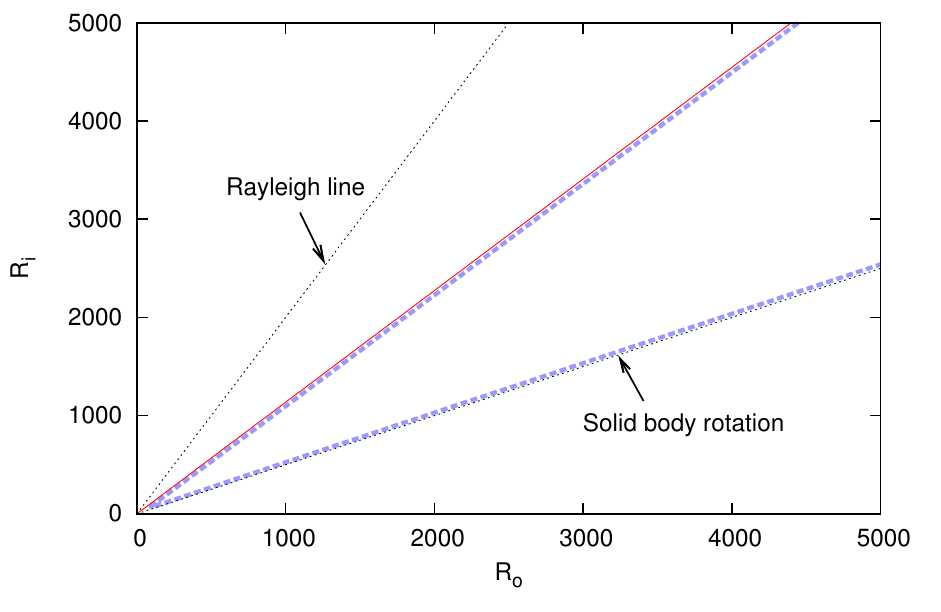} 
\caption{Upper panel: the ideal limit solutions for the perfectly conducting wide gap cylinders. Equation (\ref{Bessel}) is used. The horizontal red dashed line is the narrow-gap result $\pi^2/4$; see (\ref{ngidealpd}). 
Lower panel: the comparison of the ideal result (thin red line) and the full neutral curve for $\eta=0.5, \beta=0.2, P=1$. 
The latter result found by the viscous resistive MHD equations (\ref{potwide}) predicts that the flow is unstable within the wedge shaped domain formed by the blue dashed curve.
}
\label{fig:longwave}
\end{figure}
The ideal limit result for the perfectly conducting case can be extended to more general wide-gap Taylor-Couette flow. Neglecting all the diffusive terms from equations (\ref{potwide}) we have 
\begin{eqnarray}
\triangle \psi+\frac{4}{\beta^2r^2}\left (\frac{R_s}{R_p}+\frac{1}{r^2} \right )\psi=0,
\end{eqnarray}
where $\beta=B_0/R_p$ is the inverse magnetic Mach number (Note that this parameter is redefined from the narrow-gap case for the sake of simplicity).
The most dangerous mode occurring at $k=0$ can be found using Bessel functions of fractional order $\pm \mu$, where
\begin{eqnarray}
 \mu=\sqrt{1-\frac{4R_s}{R_p\beta^2}}.
\end{eqnarray}
The boundary conditions for $\psi$, shown in (\ref{conducting}), can be satisfied if the dispersion relation
\begin{eqnarray}
0=f(R_p/R_s,\beta,\eta)=J_{\mu}\left (\frac{2}{\beta r_i}\right )J_{-\mu}\left (\frac{2}{\beta r_o}\right )-J_{\mu}\left (\frac{2}{\beta r_o}\right )J_{-\mu}\left (\frac{2}{\beta r_i} \right )\label{Bessel}
\end{eqnarray}
holds. Here $J_{\mu}$ is the Bessel function of $\mu$th order. The behaviour of the critical value $R_s/R_p$ found by (\ref{Bessel}) is shown in the upper panel of figure 4. The rescaled variable $(R_s/R_p)(1-\eta)^2/\beta^2$ is used for the vertical coordinate because this quantity tends to $\omega/\alpha^2$ at the narrow-gap limit. 
The convergence to the narrow gap result (\ref{ngidealpd}), shown by the red dashed line, can indeed be found in the figure when $\eta$ approaches to 1. 
For $\eta<1$ the rescaled variable in the figure do not become a constant, although they tend to a constant for large $\beta$. 


The lower panel of figure 4 compares the ideal result with the full stability for $\eta=0.5, \beta=0.2, P=1$. 
The result is plotted in $R_o$--$R_i$ plane; here the entire counter-rotation regime ($R_oR_i<0$) and the region above the Rayleigh line $R_s=0$ ($R_i=\eta^{-1}R_o$) is unstable in terms of (\ref{Raycond}). 
In addition to the Rayleigh unstable regime, the MRI could also occur at the anticyclonic regime, which is the wedge shape region between the Rayleigh line and the solid body rotation line $R_p=0$ ($R_i=\eta R_o$) in the figure. 
For the selected parameters the neutral curve sits in this region as shown by the thick blue dashed curve. 
This curve was computed by solving differential equations (\ref{potwide}) with the boundary conditions (\ref{conducting}) by the Chebyshev collocation method. 

The behaviour of the upper neutral curve can be captured by the ideal result shown by the thin red line. 
For given $\eta, \beta$ the slope of the line can be found from the upper panel. From the trend of the asymptotic result, we can find that the larger the value of $\beta$ is, the more the ideal line approaches to the solid body rotation line. 
Thus, as we have seen in the upper panel of figure 2, the ideal result describes the stabilisation effect by the applied magnetic field. 
(Likewise the lower neutral curve corresponds to the high-rotation limit to be shown in figure 5. When $\beta$ is fixed, this limit coincides with the solid-body rotation line.)



\section{The long wavelength limits}

When the wavelength is large enough (equivalently $k$ gets small) we must keep some diffusive terms even at the large Reynolds number limit. This is because the driving mechanisms by the shear and the magnetic field are all multiplied by $k$ as seen in (\ref{potwide}); hence the underlying physics of the limit is somewhat similar to the Prandtl's boundary layer theory. 
Here we focus only on the narrow-gap limit equations for the sake of simplicity. The parameters $P,\alpha$ are $O(1)$ constants throughout this section.

\subsection{The shearing box computations}

We start the asymptotic analysis for the periodic boundary conditions. 
First we note that by writing
\begin{eqnarray}
K=\frac{P(kR\alpha)^2}{l^4},\qquad \Omega=\frac{\omega}{\alpha^2},\qquad \mathcal{P}=\frac{P}{1-\omega},\label{defKOP}
\end{eqnarray}
equation (\ref{comba}) can be simplified to
\begin{eqnarray}
(1+K)^2l^2=K\Omega(\mathcal{P}^{-1}+K).\label{pdlw1}
\end{eqnarray}
In the long wavelength limit we fix $(kR)$ as an $O(R^0)$ value while taking the limit of large $R$. 
At the limit equation (\ref{pdlw1}) is unchanged except that $l=n\pi/2$ becomes independent of $k$.

The quantity $K$ can be regarded as the square of the scaled axial wavenumber. 
Thus in order to find the approximation of the neutral curve we must optimise $\omega$ against $K$ for fixed $P,\alpha$. Differentiating (\ref{pdlw1}) by $K$ and requiring $\partial_K \omega=0$, we find that the extremum occurs at
\begin{eqnarray}
K=\frac{\frac{\Omega}{2 \mathcal{P}}-l^2}{l^2-\Omega}.\label{pdlw2}
\end{eqnarray}
Eliminating $K$ from (\ref{pdlw1}) and (\ref{pdlw2}) we get
\begin{eqnarray}
\Omega=4l^2\mathcal{P}(1-\mathcal{P}).\label{pdlw3}
\end{eqnarray}
We can further use it to eliminate $\Omega$ from (\ref{pdlw2}):
\begin{eqnarray}
K=\frac{1}{1-2\mathcal{P}}.\label{KPKP}
\end{eqnarray}
%

The most dangerous perturbation has $n=1$, so the neutral condition is simply
\begin{eqnarray}
\Omega=\pi^2 \mathcal{P}(1-\mathcal{P})\label{pdlw3}
\end{eqnarray}
from (\ref{pdlw3}).
The right panel of figure 5 shows the asymptotic result for $\alpha=0.1$. 
Here we only plot the curve satisfying $1-2\mathcal{P}\geq 0$ because it is inverse of $K$ from (\ref{KPKP}) and hence must be positive by definition (\ref{defKOP}). 
With increasing $P$ from very small value the flow is eventually stabilised until the sharp corner of the neutral curve between $P=0.1$ and 1. 
This corner is associated with the limit of $\mathcal{P}\rightarrow 1/2$ where the value of $K$ tends to infinite. 
On the right of the corner we have a horizontal line corresponding to the ideal result (\ref{ngidealpd}). 
The neutral points at $P=10^{-6}$ (open reverse triangle) and $1$ (open triangle) were used for the asymptotic approximations of the blue thick dashed curves in figure 1.

The left panel is the same result but for $\alpha=1$.
Unlike the previous case there are two islands of instability, which are indeed the counterparts of those seen in the full analysis. 
The horizontal line at the Rayleigh stable region is the ideal result, whilst the curve at the Rayleigh unstable region is the long wavelength limit result.
The latter instability disappears at the critical value of $P\approx 0.02675$, below which there are three neutral points. The three neutral points at $P=10^{-6}$ (filled square, diamond, and reverse triangle) and one neutral point at $P=1$ (filled triangle) are the asymptotic limits seen in the bottom and top panels of figure 1, respectively.
The behaviour of the full stability curve around the critical value of $P$ is illustrated in figure 6. The disappearance of the long-wavelength limit exactly corresponds to the disappearance of the full instability in the Rayleigh unstable regime. 

\begin{figure}
\centering
\begin{tabular}
[c]{cc}
\includegraphics[scale=1.2]{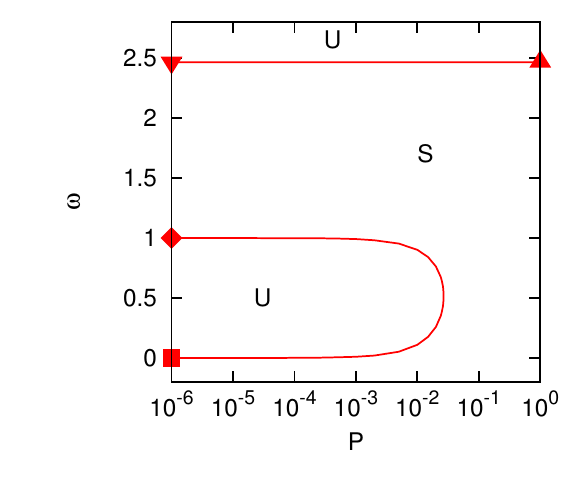} &
\includegraphics[scale=1.2]{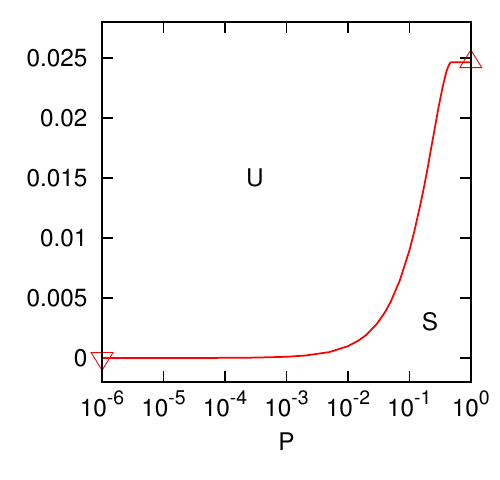}
\end{tabular}
\caption{The neutral curve at the long wavelength limit for the shearing box case. The unstable region is labeled by `U', the stable region is labeled by `S'. Left panel: $\alpha=1$, right panel: $\alpha=0.1$.}
\label{fig:longwave}
\end{figure}

\begin{figure}
\centering
\includegraphics[scale=1.2]{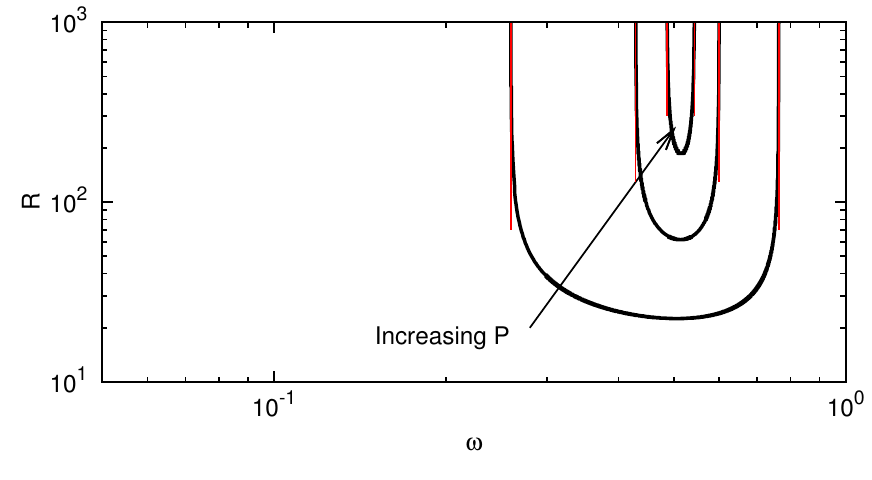} 
\caption{
The disappearance of the instability in Rayleigh unstable regime with increasing $P$.
The shearing box is used for $\alpha=1$. 
The solid curves are the full neutral curves for $P=0.02,0.026,0.0267$. 
The thin red lines are the corresponding long wave length asymptotic limits; see the left panel in figure 5.}
\label{fig:longwave}
\end{figure}


\subsection{The perfectly conducting conditions}
The similar analysis is possible for the perfectly conducting cases. 
The long wavelength limit equations can be found from (\ref{potnarrow}) straightforwardly taking the limit of $R\rightarrow \infty$ while fixing $kR$ as a constant. This is the regular limit so all the boundary conditions (\ref{ngcd}) are satisfied. 
The computational results shown in the top two panels in figure 7 analogue to those in figure 6. 
Similar to the previous figure, the symbols on the curves represent the corresponding asymptotic results shown in figure 2. 

Most of the neutral perturbations are steady, although oscillatory neutral modes are detected for some values of $\alpha$ as shown by the green dashed curve in the right panel of figure 7. 
The behaviour of the steady modes are similar to the shearing box computation, because setting $\sigma=0$ the long wavelength equations can be combined into the single equation
\begin{eqnarray}
\{\partial_y^4+(P^{1/2} \alpha R k)^2\}^2\psi'' 
+(P^{1/2}\alpha R k)^2 \frac{\Omega}{\mathcal{P}} \psi''''+\Omega (P^{1/2}\alpha R k)^4 \psi=0. 
\end{eqnarray}
This equation suggests that the steady stability result obtained by optimising against the scaled wavenumber $(P^{1/2}\alpha R k)$ can be summarised by using $\Omega$ and $\mathcal{P}$ defined in (\ref{defKOP}). 
The lower panel of figure 7 is the long wavelength results in terms of $\Omega=\omega/\alpha^2$ and $\mathcal{P}=P/(1-\omega)$. 
For any choice of $\alpha$, the steady modes are on the red curves in the figure (The horizontal line is the ideal result appearing when $P^{1/2}\alpha R k\rightarrow \infty$).
The oscillatory mode appears only when the value of $\alpha$ is smaller than $\approx 0.2$ as shown by the green dashed curves, which are indeed out of the red curve. 
We also remark that the oscillatory mode can also be found at finite Reynolds numbers.
For example, on the thick blue dashed curve in figure 2 such mode occurs between the blue circles around the Rayleigh line.


Another minor qualitative difference to the shearing box computation is that the instability in the Rayleigh unstable region exists for a while beyond the critical value of $P$ found by the asymptotic analysis. 
Figure 8 shows the behaviour of the full neutral curve around the critical value $P\approx 0.00248$ for $\alpha=1$ found in the top left panel of figure 7. 
There is a remnant of the instability enclosed by the neutral curve unlike figure 6, but it eventually shrinks with increasing $P$ and vanishes at $P\approx 0.00641$.

\begin{figure}
\centering
\begin{tabular}
[c]{cc}
\includegraphics[scale=1.2]{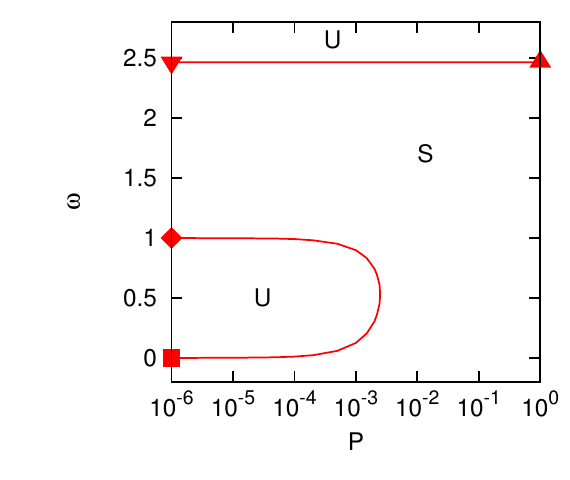} &
\includegraphics[scale=1.2]{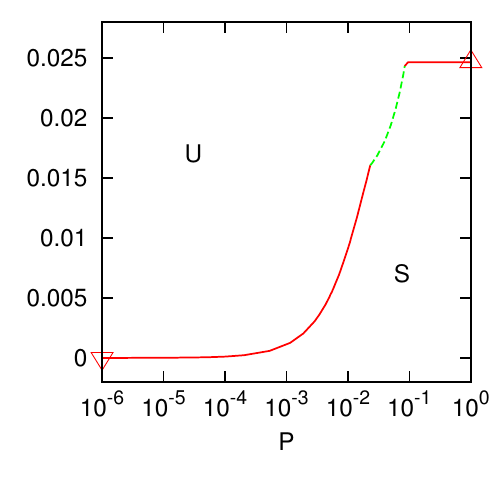}
\end{tabular}\\
\includegraphics[scale=1.2]{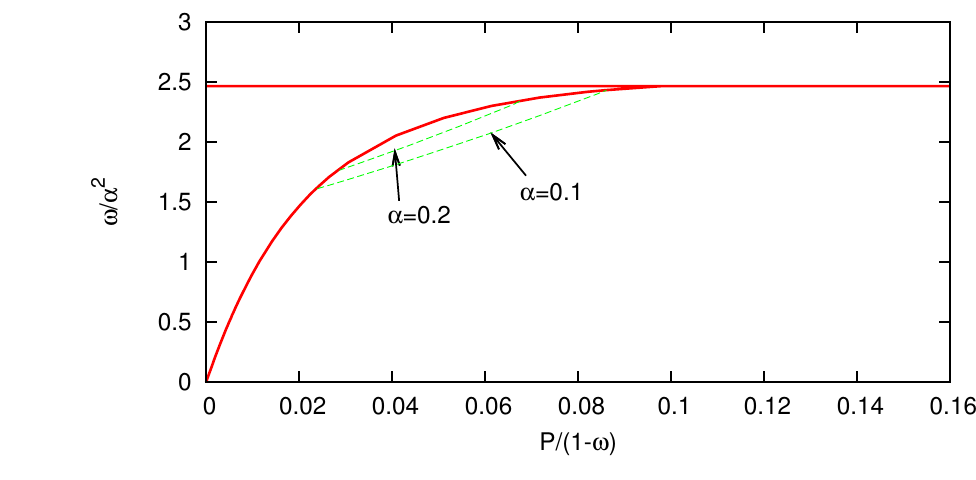} 
\caption{The top two panels are the same results as figure 4 but for the perfectly conducting boundary conditions. 
The red solid/green dashed curves are the steady/oscillatory modes. 
The bottom panel is the results summarised in terms of  $\omega/\alpha^2$ and $P/(1-\omega)$.
The red solid curve is the neutral steady modes for various $\alpha$. The green curves are the neutral oscillatory modes for $\alpha=0.1$ and 0.2.}
\label{fig:longwave}
\end{figure}


\begin{figure}
\centering
\includegraphics[scale=1.2]{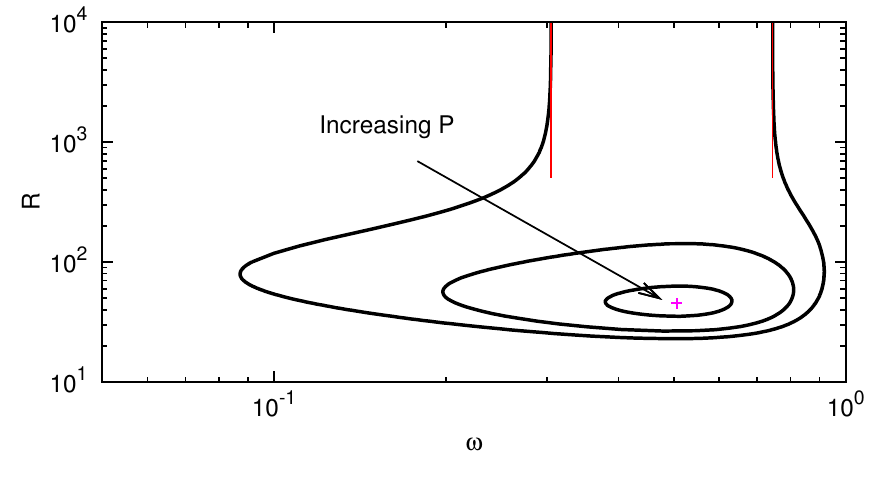} 
\caption{The same results as figure 5 but for the perfectly conducting boundary conditions. 
The solid curves are the full neutral curves for the $P=0.002,0.004,0.006$, where equations (\ref{potnarrow}) are solved for $\alpha=1$.
The unstable region shrinks with increasing $P$ and disappear at the cross point when $P=0.00641$. 
The red lines are the long wavelength asymptotic result for $P=0.002$; see the top left panel of figure 7.}
\label{fig:longwave}
\end{figure}

\subsection{The perfectly insulating conditions}

As we have seen in figure 3 the asymptotic behaviour of the left hand branch of the perfectly insulating case is quite different from the other two boundary conditions. Nevertheless, the asymptotic limit can be explained by the long wavelength limit $kR\sim O(R^0)$ but with fixed $\omega R$.


In order to find the limiting reduced system it is convenient to choose $k$ as a perturbation parameter rather than $R^{-1}$. Let us define the rescaled $O(k^0)$ parameters $\widetilde{k}=P^{1/2} \alpha k R$, $\widetilde{\omega}=P^{1/2}R\omega/\alpha$ and expand
\begin{eqnarray}
\psi=\psi_0+k\psi_1+\cdots,\qquad 
P^{1/2}R\omega k v=v_0+kv_1+\cdots, \\
R\omega k b=b_0+kb_0+\cdots,\qquad
P^{1/2}\phi=\phi_0+k\phi_1+\cdots.
\end{eqnarray}
Substituting those expansions to (\ref{potnarrow}), at $O(k^0)$ we can find
\begin{subequations}
\begin{eqnarray}
 \phi_0'''' +i\widehat{k} \psi_0''+iv_0=0,\qquad
v_0''+i\widehat{k}b_0=0,\\
 \psi_0''+i\widehat{k}\phi_0=0,\qquad
 b_0''+i\widehat{k}v_0=0.
\end{eqnarray}
\end{subequations}
Note that $\psi_0'=\phi_0=0$ must be satisfied at $x=\pm 1$.
This problem is exactly what we saw in the ideal limit at the optimum $\omega=0$. 
Within the asymptotic framework we interpret this as the vanishing leading order terms of $O(R^0)$ in the asymptotic expansion of $\omega$.
Therefore we must look the higher order term in the expansion, namely $\widetilde{\omega}$, to determine the stability.

In order to fix the value of $\widetilde{\omega}$ we must look the next order equations at $O(k)$. 
Using the leading order solutions $\psi_0=1, \phi_0=0$ we have
\begin{subequations}\label{1inslw}
\begin{eqnarray}
 \phi_1'''' +i\widetilde{k} \psi_1''+iv_1=0,\qquad
v_1''+i\widetilde{k}b_1=0,\\
 \psi_1''+i\widetilde{k}\phi_1=0,\qquad
 b_1''+i\widetilde{k}v_1=i\widetilde{k}\widetilde{\omega},
\end{eqnarray}
\end{subequations}
which should be solved together with the higher-order boundary conditions
\begin{eqnarray}
v_1=\phi_1=\phi_1'=b_1=0, ~~~~\psi_1'\pm 1=0 ~~~~\text{at}~~~~ x=\pm 1.\label{bdinslw}
\end{eqnarray}
This is inhomogeneous form of the leading order equations so the solvability condition gives the dispersion relation.

In order to solve (\ref{1inslw}) we combine the equations to find that the quantity $\Psi=\psi_1''+\widetilde{\omega}/\widetilde{k}$ satisfies the simple differential equation
\begin{eqnarray}
\frac{\Psi^{(8)}}{\widetilde{k}^4}+2\frac{\Psi^{(4)}}{\widetilde{k}^2}+\Psi=0.\label{eqpsipsi}
\end{eqnarray}
Assuming $\Psi$ is an even function, the general solution of (\ref{eqpsipsi}) can be found as $\Psi=A_1S_S+A_2C_C+A_3\kappa xS_C+A_4 \kappa xC_S$, where
\begin{eqnarray}
S_S=\sin (\kappa x)\sinh(\kappa x), \qquad C_C=\cos (\kappa x)\cosh(\kappa x),\\
S_C=\sin (\kappa x)\cosh(\kappa x), \qquad C_S=\cos (\kappa x)\sinh(\kappa x), 
\end{eqnarray} 
and $\kappa=(\widetilde{k}/2)^{1/2}$.
The five unknown constants $A_1,A_2,A_3,A_4$ and $\widetilde{\omega}$ can be fixed by applying the following five conditions found from (\ref{1inslw}) and (\ref{bdinslw}):
\begin{eqnarray}
\Psi^{(4)}=\Psi^{(1)}=\Psi^{(6)}+\widetilde{k}^2\Psi^{(2)}=0,\qquad \Psi=\frac{\widetilde{\omega}}{\widetilde{k}},\qquad \text{at} \qquad x=1
\end{eqnarray}
and
\begin{eqnarray}
\int^1_{0} \Psi dx=-1+\frac{\widetilde{\omega}}{\widetilde{k}}.
\end{eqnarray}
After some algebra the dispersion relation can be found as 
\begin{eqnarray}
\widetilde{\omega}(\widetilde{k})=\frac{\widetilde{k}\kappa \gamma_3}{(\kappa \gamma_1+\lambda_1) g_1+(\kappa \gamma_2+\lambda_2) g_2+(\kappa \gamma_3+\lambda_3)},\label{disp}
\end{eqnarray}
where 
\begin{eqnarray}
\gamma_1=\{S_C(C_S^2+S_C^2-S_S^2+C_C^2)-2C_S S_S C_C\},\\
\lambda_1=\{(C_C-S_S)(S_C^2-C_S^2)/2-(S_S+C_C)C_S S_C\},\\
\gamma_2=\{C_S(C_S^2+S_C^2+S_S^2-C_C^2)-2S_C S_S C_C\},\\
\lambda_2=\{(C_C+S_S)(S_C^2-C_S^2)/2-(S_S-C_C)C_S S_C\},\\
\gamma_3=\{S_S(S_C-C_S)+C_C(S_C+C_S)\}, \qquad \lambda_3=-(C_S^2+S_C^2),\\
g_1=-\frac{S_S-C_C}{4(S_S^2+C_C^2)},\qquad
g_2=-\frac{S_S+C_C}{4(S_S^2+C_C^2)}.
\end{eqnarray}


The minimum of $\widetilde{\omega}(\widetilde{k})$ can be found from (\ref{disp}) as $17.45$ at $\widetilde{k}= 7.006$. This gives the asymptotic result
\begin{eqnarray}
R=\frac{17.45\alpha}{\omega P^{1/2}}.\label{inlong}
\end{eqnarray}
The asymptotic lines in figure 3 excellently captures the large Reynolds number behaviour of the full neutral curve.

\section{The high rotation limits}

\subsection{The shearing box computations}

In order to find the large $\omega$ asymptotic limits seen in figures 1, 2 and 3, it is convenient to use the normalised form similar to (\ref{scaled}). We shall see that the limiting neutral curve is governed by the non-dimensional parameter
\begin{eqnarray}
C=\frac{S^2}{\omega},\label{defCC}
\end{eqnarray}
which is essentially the Rossby number multiplied by square of the Lundquist number $S=R \alpha P$. 

The narrow-gap analog of (\ref{scaled}) can be found by writing $V=\alpha v, \Phi=PR\alpha \phi, B=P^{-1}R^{-1}b$ in equations (\ref{potnarrow}). 
\begin{subequations}\label{horescale}
\begin{eqnarray}
\frac{\triangle^2 \Phi}{R^4P^3\alpha^2} +ik \frac{\triangle \psi}{R^2P^2}+ik\frac{V}{C}=0,\label{hom1}\\
\frac{\triangle V}{R^2P\alpha^2}+ik B-ik\left (\frac{1}{C}-\frac{1}{R^2P^2\alpha^{2}} \right )\Phi=0,\label{hom2}\\
\triangle \psi+ik \Phi=0,\\
\triangle B+ik V-ik\psi=0.
\end{eqnarray}
\end{subequations}
We take the large $R$ limit of this system fixing the rescaled variables and $k$, $\alpha$, $P$, $C$ as $O(R^0)$ quantities. 
From those assumptions and (\ref{defCC}) we see that $\omega$ is indeed asymptotically large. 

From (\ref{hom1}), (\ref{hom2}) the hydrodynamic part of the solution is simply
\begin{eqnarray}
V=0, \qquad \Phi=CB.
\end{eqnarray}
Then using them to the other two equations we have the eigenvalue problem for $C$
\begin{eqnarray}
\triangle^2\psi-Ck^2 \psi=0.\label{highrocore}
\end{eqnarray}
As we have seen in the previous sections the sinusoidal function of wavelength $4/n, n\in \Natural$ can be used to derive the dispersion relation
\begin{eqnarray}
C(k,n)=\frac{((n\pi/2)^2+k^2)^2}{k^2}.
\end{eqnarray}
The minimum $C=\pi^2\approx 9.870$ found at $k=\pi/2, n=1$ gives the high-rotation limit of the neutral curve \begin{eqnarray}
R=\frac{\sqrt{\pi^2\omega}}{\alpha P},\label{hosb}
\end{eqnarray}
which gave the excellent asymptotic prediction in figure 1.

The high-rotation limit line (\ref{hosb}) describes the transition from the hydrodynamic instability to the MRI at the anticyclonic regime. The key parameter is the Lundquist number as pointed out in Kirillov \& Stefani (2011) who used some special property of the algebraic equation (\ref{comba}) rather than the asymptotic approach. From figure 1 we can see that for given $\omega$:
\vspace{2mm}
\\
(i) If $R\ll O(1/\alpha P)$, namely the Lundquist number is small, 
then the flows satisfying Rayleigh's condition (\ref{Raycond}) are stable,\\
(ii) If $R\gg O(1/\alpha P)$, namely the Lundquist number is large, then 
the flows satisfying the Velikhov-Chandrasekhar condition (\ref{VCcond}) are stable.
\vspace{2mm}

We have remarked earlier that the stability must approach to that for purely hydrodynamic case when at least either $P$ or $\alpha$ gets smaller. The Velikhov-Chandrasekhar paradox occurs because in this argument we have implicitly assumed that $R$ is $O(1)$ while taking the limit. For large enough $R$, the limiting result (\ref{hosb}) suggests that the MRI must appear in the Rayleigh stable region no matter how small $P$ and $\alpha$ are, as long as they are finite. 
This conclusion is the same as Kirillov \& Stefani (2011) overall, but what is remarkable here is that our asymptotic approach can be extended to more general cases, as we shall see in the subsequent subsections. 

For small $P$ the magnetic Reynolds number $R_m=PR$ becomes also important in view of Goodman \& Ji (2002). 
Actually our high rotation asymptotic approximation is valid only when $R_m$ is sufficiently large. 
In order to see this consider the critical value of $R_m$ above which we can find an appropriate size of magnetic field to switch on the MRI; this is equivalent to draw the envelope of the neutral curve for various values of $\alpha$. 
The black solid curve in figure 9 shows the envelope for $P=10^{-6}$. For such small magnetic Prandtl number we can see clear separation of hydrodynamic instability and the MRI, where the latter only occurs when $R_m$ is large.
Any neutral curve with fixed $\alpha$ should sit above this envelope; see the thin grey curve computed for $\alpha=1$.

%

We can approximate the critical value of $R_m$ from the high-rotation limit of the envelope. 
Let us rewrite (\ref{potnarrow}) using $V=\alpha v$, $\varphi=\alpha \phi$, $R_m=PR$, $A=\alpha^2/\omega$ to have the rescaled system
\begin{subequations}\label{scaledAomega}
\begin{eqnarray}
\omega^{-1}A^{-1}\triangle^2 \varphi +ik R (\triangle \psi+A^{-1}V)=0,\\
\omega^{-1}A^{-1}\triangle V+ik R( b-(1-\omega^{-1})A^{-1}\varphi )=0,\\
\triangle \psi+ik R_m\varphi=0,\\
\triangle b+ik R_m V-ikR_m\psi=0.
\end{eqnarray}
\end{subequations}
If we consider the large $\omega$ limit assuming $k,V,\phi,b,\psi,R_m,A$ are $O(\omega^0)$, then equations (\ref{scaledAomega}) can be reduced to 
\begin{eqnarray}
\triangle^2 \psi-k^2 R_m^2A(A\triangle \psi+\psi)=0.
\end{eqnarray}
The use of the sinusoidal function yields the dispersion relation
\begin{eqnarray}
R_m^{2}
=\frac{l^4}{k^2(A-l^2A^2)}.
\end{eqnarray}
We must minimise the right hand side against the wavenumbers $k,n$ and the normalised magnetic field $A$ to find the approximation of the envelope. 
The optimised value 
\begin{eqnarray}
R_m=\sqrt{27}\frac{\pi^2}{4}\approx 12.821\label{Rmpd}
\end{eqnarray}
at $k=\pi/2\sqrt{2}\approx 1.111$, $n=1$, $A=1/2l^2\approx 0.1351$ is the horizontal dotted line in figure 9.
If $R_m$ is larger than this critical value, we can expect that (\ref{hosb}) approximates the asymptotic behaviour of the MRI.


\begin{figure}
\centering
\includegraphics[scale=1.2]{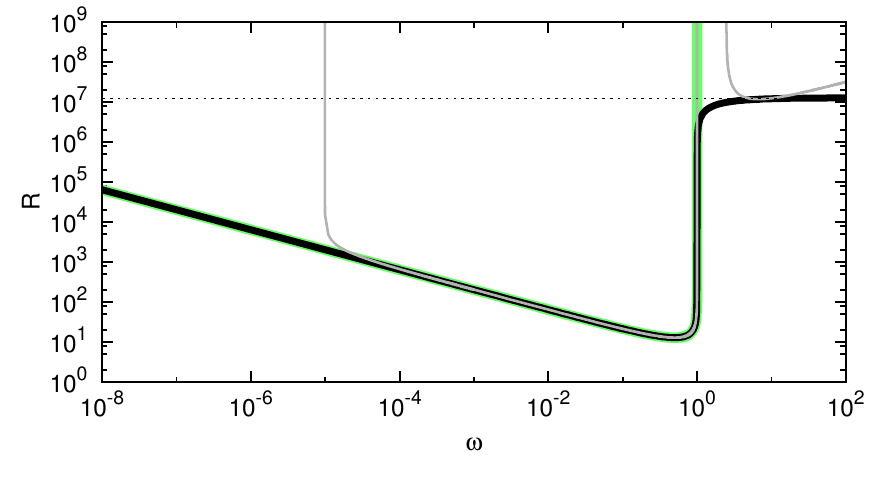} 
\caption{The thick black curve is the envelope of the neutral curves for various $\alpha$. 
The periodic boundary conditions are used and $P=10^{-6}$. 
The thin grey curve is the neutral curve for $\alpha=1$, taken from figure 1. 
The thick green curve is the purely hydrodynamic result ($\alpha=0$).
The horizontal dotted line is the high-rotation asymptotic result for the MRI (\ref{Rmpd}). 
}
\label{fig:longwave}
\end{figure}

\subsection{The narrow-gap limit results}

Now let us include the effect of the walls. 
Since the viscous terms drop in the high-rotation limit we must consider near wall boundary layers to satisfy the boundary conditions for the hydrodynamic parts. 
That asymptotic structure is similar to the small $P$ analysis by Goodman \& Ji (2002), but again we insist that we are concerning the different asymptotic regime.


First we consider the perfectly conducting walls.
The limiting neutral condition can again be found by (\ref{highrocore}) but now we must apply the boundary conditions
\begin{eqnarray}
\psi =\psi'''-k^2\psi'=0 \qquad \text{at}\qquad x=\pm 1
\end{eqnarray}
that ensure the magnetic boundary conditions $\psi=B'=0$ on the walls.
The boundary conditions for the hydrodynamic parts must be taken care of by the near wall boundary layer of thickness $O(R^{-1})$. 
The boundary layer equations near $x=\pm 1$ can be obtained by simply considering $\Phi,V,\psi,B$ as a function of the stretched variable $X=(x\mp 1)/R$ in (\ref{horescale}). 
Within the boundary layer $\psi$ and $B$ can be written by local Taylor expansion of outer solutions, and hence they are merely a linear function and a constant, respectively. 
The rest of the solutions $V,\Phi$ can be found by the rescaled hydrodynamic equations
\begin{eqnarray}
\frac{\Phi_{XXXX}}{P^3\alpha^2}+ikC^{-1} V=0,\\
\frac{V_{XX}}{P\alpha^2}+ik B-ik C^{-1}\Phi=0,
\end{eqnarray}
where $V\rightarrow 0,\Phi\rightarrow CB$ as $|X|\rightarrow \infty$.
There are three decaying and three growing roots in the characteristic equation so the inner solution matching to the outer solution can satisfy all the required boundary conditions on the walls. 
Clearly the boundary layer is passive, in the sense that it does not affect to the outer eigenvalue problem.

The minimum of $C$ occurs at the limit of $k\rightarrow 0$, and that limiting value can be found by substituting the expansion
\begin{eqnarray}
\psi=\psi_0+k^2\psi_1+\cdots
\end{eqnarray}
to the outer equation (\ref{highrocore}).
The leading order problem is 
\begin{eqnarray}
\psi_0''''=0
\end{eqnarray}
with the boundary conditions
$\psi_0=\psi_0'''=0$ at $x=\pm 1$.
Using the solution $\psi_0=1-x^2$ to the next order problem, we have the inhomogeneous problem
\begin{eqnarray}
\psi_1''''-2\psi_0''-C\psi_0=0 \label{firstho}
\end{eqnarray}
subject to $\psi_1=\psi_1'''-\psi_0'=0$ at $x=\pm 1$.
We can integrate (\ref{firstho}) once to show $C=3$. Therefore, the high-rotation limit for the perfectly conducting boundary conditions can be found as
\begin{eqnarray}
R=\frac{\sqrt{3\omega}}{\alpha P}.\label{cdholine}
\end{eqnarray}

For the perfectly insulating case we must solve (\ref{highrocore}) together with the magnetic boundary conditions $\psi'\pm k\psi=0$ and $\psi''- k^2\psi=0$ (The hydrodynamic conditions are again satisfied through the boundary layer solutions similar to the perfectly conducting case). 
The general solutions can be found by exponential functions and the boundary conditions are satisfied if the dispersion relation
\begin{eqnarray}
f(C,k)=l_+(\tanh l_+)-l_-(\tan l_-)+2k=0, \qquad l_{\pm}=\sqrt{k\sqrt{C}\pm k^2}
\end{eqnarray}
holds. A little numerical work yields the minimum $C=6.598$ at $k=1.029$. 
Thus the high-rotation limit for the perfectly insulating boundary walls is
\begin{eqnarray}
R=\frac{\sqrt{6.598\omega}}{\alpha P}.\label{inholine}
\end{eqnarray}

Both (\ref{cdholine}) and (\ref{inholine}) excellently predict the behaviour of the full numerical neutral curves shown in figures 2 and 3, respectively.
Similar to the shearing box case, those approximations are valid when $R_m$ is sufficiently large.
The critical values of $R_m$ similar to (\ref{Rmpd}) can be derived for the wall bounded cases as derived in Appendix B and summarised in table 1.

\subsection{Wide gap cases}

\begin{figure}
\centering
\includegraphics[scale=1.2]{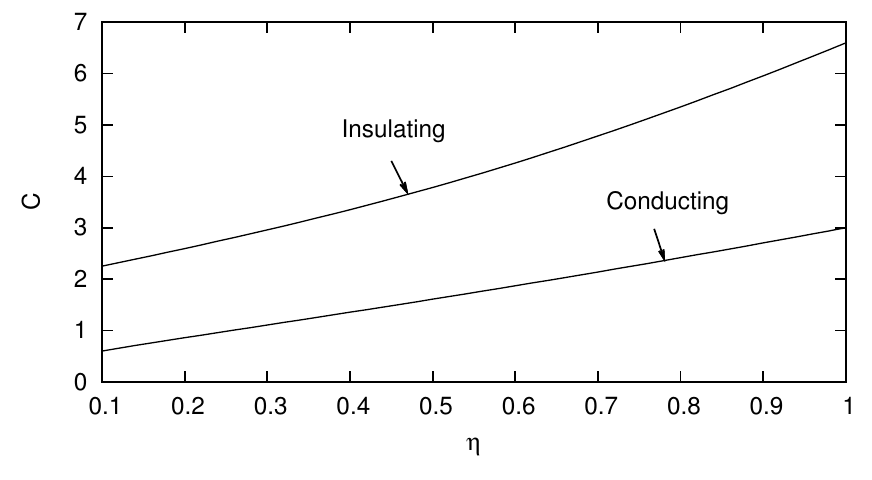} \\
\includegraphics[scale=1.2]{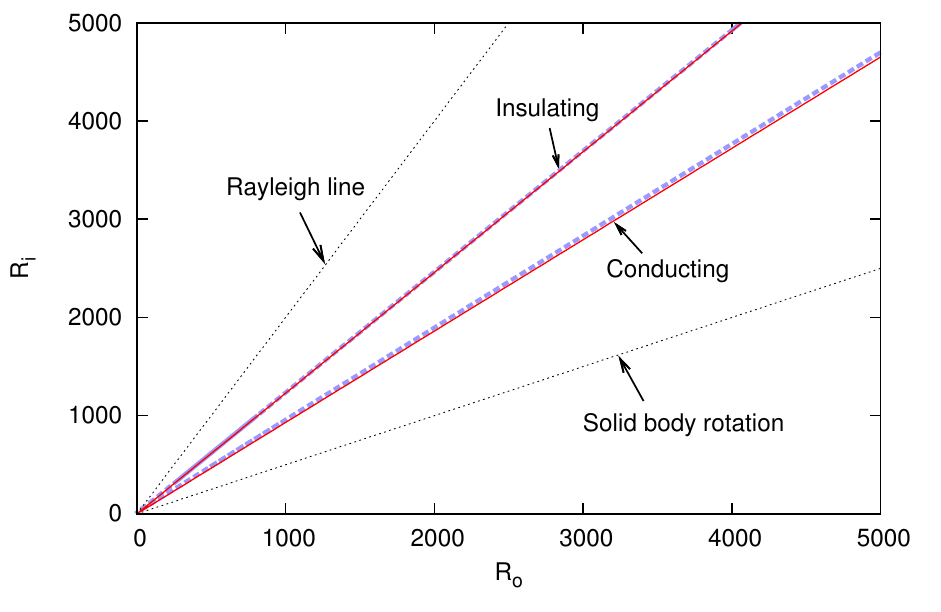} 
\caption{
Upper panel: The solution of the asymptotic problem (\ref{asywideho}).
Lower panel: The neutral curves for $\eta=0.5$, $B_0=2$, $P=1$. 
There are two thick blue dashed neutral curves found by applying two boundary conditions (\ref{conducting}), (\ref{insulating}) for the full viscous resistive MHD equations (\ref{potwide}). 
Above those curves the corresponding flows are unstable. 
The two red solid lines are the asymptotic result (\ref{criticalHR}) with the values of $C$ taken from the upper panel.}
\label{fig:longwave}
\end{figure}


Finally we shall extend the above high-rotation limits to the wide gap cases. 
Here we fix $B_0$ rather than the magnetic Mach number; in fact most of the previous numerical computations concerned constant magnetic field cases. 
This does not cause any major change of the asymptotic structure discussed earlier. 
Thus the outer region equations can be found by taking $R_p\rightarrow \infty$ in the rescaled system (\ref{scaled}). 

From the hydrodynamic part of the equations, (\ref{moscaled1}), (\ref{moscaled1}), we find simple solutions $V=0$ and $B=2\Phi/Cr_o^2$. 
Further using them to (\ref{indscaled1}), (\ref{indscaled2}) we have the eigenvalue problem
\begin{subequations}\label{asywideho}
\begin{eqnarray}
\triangle^2 \psi -Ck^2 \frac{r_o^2}{r^2}\psi=0,
\end{eqnarray}
\end{subequations}
which must be solved subject to the magnetic part of the boundary conditions, namely (\ref{conducting}) or (\ref{insulating}) replacing $b$ by $\triangle \psi$.
Again thin near wall boundary layers must be inserted to satisfy the hydrodynamic part of the boundary conditions, but we omit further analysis because they are passive. 
Given $\eta$ we can numerically find the minimum of the eigenvalue $C$ against $k$; the results are summarised in the upper panel of figure 10. 

From the definitions of $C$, $R_p$ and $R_s$ (see (\ref{scaledV}), (\ref{defRsRp})), we arrive at the conclusion that the neutral curve in the $R_i$-$R_o$ plane asymptotes to the straight line
\begin{eqnarray}
R_i=\frac{C+S^2\eta^2}{\eta(C+S^2)}R_o\label{criticalHR}
\end{eqnarray}
which sits between the Rayleigh line and solid body rotation line. 
The lower panel of figure 10 compares that asymptotic result with the full numerical stability result, and we find excellent agreements for both of the boundary conditions. 
(Similar to the previous cases, when $P$ is small such agreement could be found only when the magnetic Reynolds number is sufficiently larger than the critical value; see Appendix B.)
The slope of the asymptotic line is controlled solely by the scaled Lundquist number $C^{-1/2}S$. 
With increasing that parameter the anticyclonic region is eventually destabilised. 
In particular, we have the limiting behaviours of the slope
\begin{subequations}\label{limitHR}
\begin{eqnarray}
R_i/R_o&\rightarrow& \eta^{-1} ~~\text{(The Rayleigh line)~~~~~~~~~~~~~~~~~~as~~~~} C^{-1/2}S \rightarrow 0,\\
R_i/R_o&\rightarrow& \eta ~~~~~\text{(The solid body rotation line)~~~~as~~~~} C^{-1/2}S \rightarrow \infty,
\end{eqnarray}
\end{subequations}
which exactly correspond to the stability bounds predicted by the Rayleigh's condition (\ref{Raycond}) and the Velikhov-Chandrasekhar condition (\ref{VCcond}), respectively.
The coefficient $C$ carries information of the boundary conditions to the asymptotic stability result.
From the upper panel of figure 10 the conducting case is always more unstable than the insulating case. 





\section{Conclusion}

We have performed large Reynolds number asymptotic analyses for the neutral curve of the magnetised Taylor-Couette flow, assuming finite magnetic Prandtl number. The local analysis or equivalently shearing box computation is first used to predict qualitative parameter dependence of the stability, and then the result is extended to include the effect of walls and the curvature. 
The asymptotic results are compared with the numerical solutions of the linearised MHD equations with the viscosity and resistivity fully retained. 

In order to focus on the effect of the walls, we began the calculation by the narrow gap cases defining the shear Reynolds number $R$ and the inverse Rossby number $\omega$.  
The neutral curves obtained by imposing periodic conditions, no-slip perfectly conducting conditions, and no-slip perfectly insulating conditions were plotted in figures 1,2 and 3, respectively.
When the magnetic Prandtl number is not too small, there is one neutral curve describing the MRI mode as shown in the upper panels. 
The behaviour of the left hand branch can be explained by the ideal result, whilst the right hand branch tends to what we referred to as the high-rotation limit. For smaller magnetic Prandtl number, we can find from the lower panels that the behaviour of the neutral curve is more complicated because there is also hydrodynamic mode of the instability emerging in the Rayleigh unstable regime $\omega \in [0,1]$. 
More precisely, the large $R$ asymptotic limit of the mode can be described by the long wavelength limit derived in section 4.
The richness of the asymptotic limits suggests that the ideal limit is not sufficient to describe all the large $R$ limiting behaviours of the neutral curve.

Actually, although the ideal approximation is used in the early years of the MRI studies, the omission of the diffusivities causes some contradictions to the full analysis. 
According to the ideal result for the shearing box computation, there is a critical value of $\omega$ below which the flow is predicted to be stable. 
The stronger the external magnetic field is, the larger the critical value of $\omega$ should be, independent of the magnetic Prandtl number. 
This presents the first inconsistency to the purely hydrodynamic result at zero magnetic Prandtl number, where the instability should occur when $\omega \in [0,1]$.
Another paradox can be found when one notice that the hydrodynamic result should also be recovered for vanishing applied magnetic field. 
The ideal result predicts instability for $\omega\geq 0$ in this case, although anticyclonic regime $\omega>1$ must be stabilised - this is the narrow gap manifestation of the Velikhov-Chandrasekhar paradox.

Those curious inconsistencies occurred because in the above discussions we did not consider the size of the variables and parameters in detail. 
The asymptotic approach accounts for all delicate balances of each term, thereby resolving the paradoxical results. 
The former paradox can be overcome by the finding that with decreasing the magnetic Prandtl number the hydrodynamic mode abruptly appears at certain critical value of $P$. 
In order to find that critical value we must take asymptotically long wavelength of $O(R^{-1})$. With this scaling the diffusive terms must be retained and this is the reason why the corresponding instability cannot be predicted by the ideal result. 
%
Also the latter Velikhov-Chandrasekhar paradox can be explained by the behaviour of the high-rotation limit, where the resistivity remains the leading order effect. 
The limiting stability is solely determined by the new non-dimensional parameter made of the Rossby number multiplied by the square of the Lundquist number. 
Therefore, with increasing the Lundquist number $S$ the anticyclonic region $\omega>1$ is destabilised gradually. 
This conclusion is certainly consistent to Kirillov \& Stefani (2011).

From figures 1 and 2 we can find that the introduction of the perfectly conducting walls only cause some minor qualitative differences to the stability property; see the remark in section 4. 
The derivation of the ideal, long wavelength and high-rotation limit in this case remains similar to the shearing box case, although for some cases we must insert passive near-wall boundary layers. 
On the other hand, when the perfectly insulating walls are used there is an important difference that the ideal limit no more serve an accurate prediction of the neutral curve as seen figure 3.
The asymptotic analysis revealed that to leading order the ideal result gives rather trivial result, and hence we must look higher order.



The extension of those narrow gap results to the wide gap cases is possible. 
The ideal limit above predicts that there is a critical Rossby number for given magnetic Mach number. 
For the wide gap case the ratio $R_i/R_o$ controls the Rossby number of the flow and hence in $R_o$--$R_i$ plane we can draw the critical line whose slope depends on the magnetic Mach number; see figure 4. 
The slope of the line becomes gentler with increasing the applied magnetic field. 
The approximation is valid for the upper neutral branch, so again we have the stabilisation effect by the magnetic field exemplified by Velikhov (1959). 
Likewise the lower neutral branch in the figure can be approximated by the high-rotation limit, which represents the destabilisation effect by the applied magnetic field. 
The slope of the corresponding critical line is determined by the Lundquist number, as suggested by the narrow gap analysis. 
The simple analytic form (\ref{criticalHR}) of the line, whose limiting cases (\ref{limitHR}) connect the two ideal limits (\ref{VCcond}) and (\ref{Raycond}), indeed give an excellent prediction of the high-speed MRI stability boundary, as seen in figure 10.


Throughout the paper axisymmetry of the perturbations is assumed. 
We note in passing that non-axisymmetric perturbations are actually not significant for the long-wavelength and high-rotation limits because in order to balance the azimuthal derivative we must choose the azimuthal wavenumber $m$ to be $O(R^{-1})$. 
However, for the ideal limit we can take integer $m$ so non-axisymmetric mode may be possible. 
In fact, R\"udiger et al. (2018) reported that for some Rayleigh unstable flows the $m=1$ mode becomes most unstable when the cylinders are made of perfectly conducting walls. 
It should also be noted that the non-axisymmetry of the perturbations typically appears when the cylinders are counter rotating as widely recognised in purely hydrodynamic Taylor-Couette flow studies. 
Hence it is of interest to extend the recent asymptotic work by Deguchi (2016) to magneto-hydrodynamic flows in the future.

Support by Australian Research Council Discovery Early Career Researcher Award DE170100171 is gratefully acknowledged.

\appendix

\section{Numerical boundary conditions}
In order to numerically solve the differential equations (e.g. (\ref{potwide}), (\ref{potnarrow})) we use Chebyshev expansions modified so that they satisfy the boundary conditions automatically. 
For the hydrodynamic parts we can use the expansions
\begin{eqnarray}
\phi=\sum_{l=0}^{L} (1-x^2)^2 T_l(x),\qquad v=\sum_{l=0}^{L}(1-x^2) T_l(x)
\end{eqnarray}
which ensure the no-slip conditions $\phi'=\phi=v=0$ on the walls. Here $T_l$ is $l$th Chebyshev polynomials and the expansion is truncated at $L$th mode for the numerical purpose. 

The expansions for the magnetic parts are more tricky. The corresponding boundary conditions are written in the form of Robin's conditions
\begin{eqnarray}
g'+M_{\pm}g=0\qquad \text{at} \qquad x=\pm1\label{Robin}
\end{eqnarray}
with some constants $M_+$ and $M_-$. 
For the function $g(x)$ we can consider the expansion like
\begin{eqnarray}
g(x)=\sum_{l=0}^{L} (1-x^2)T_l(x)+\chi_0+\chi_1x
\end{eqnarray}
where each basis function satisfies (\ref{Robin}) when the constants $\chi_0$, $\chi_1$ are chosen as
\begin{subequations}
\begin{eqnarray}
\chi_0=2\frac{(-1)^l(1+M_+)+(1-M_-)}{(1-M_-)M_+ -(1+M_+)M_-},\\
\chi_1=-2\frac{(-1)^l M_+ +M_-}{(1-M_-)M_+ -(1+M_+)M_-}.
\end{eqnarray}
\end{subequations}


After substituting those expansions to the differential equations, we evaluate them at the collocation points
\begin{eqnarray}
x=\cos \left (\frac{l+1}{L+2}\pi \right), \qquad l=0,\dots,L
\end{eqnarray}
to get the algebraic equations used in the computations.

\section{Optimised magnetic field}
\begin{figure}
\centering
\includegraphics[scale=1.2]{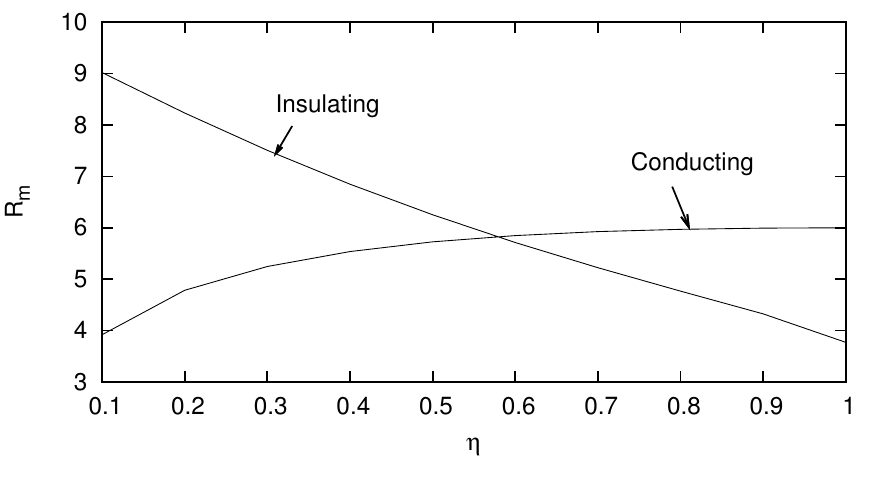} 
\caption{The critical $R_m$ similar to (\ref{Rmpd}) but for wide gap Taylor Couette flow.}
\label{fig:longwave}
\end{figure}

Here we shall find the critical value of $R_m$ to observe the MRI for the wide-gap cases, using the high-rotation limit. 
If we rewrite (\ref{potwide}) using $V=\beta v$, $\varPhi=\beta \phi$, $R_m=P(2R_p/r_m^2)$, $A=r_m^2\beta^2 R_p/4R_s$, then
\begin{subequations}
\begin{eqnarray}
\beta^{-2}\triangle^2 \varphi +ik R_p \triangle \psi+ik R_p \left ( \frac{r_m^2}{2A}+\frac{2}{r^{2}\beta^{2}} \right ) V=0,\\
\beta^{-2}\triangle V+ik R_p b-ikR_p\frac{r_m^2}{2A}\varphi=0,\\
\triangle \psi+ik R_m \frac{r_m^2}{2}\varphi=0,\\
\triangle b+ik R_m\frac{r_m^2}{2}V-ikR_m\frac{r_m^2}{r^2}\psi=0.
\end{eqnarray}
\end{subequations}
Taking the limit of large $\beta$, $R_s/R_p$, we can find the approximation of the critical value $R_m$ similar to the shearing box computation (see figure 9).
If we keep $R_m$, $A$, and all the rescaled variables to be $O(1)$, the limiting equations are
\begin{subequations}\label{AAAA}
\begin{eqnarray}
\triangle \psi+ik R_m Ab=0,\\
\triangle b-ik R_m\left (A\triangle \psi+\frac{r_m^2}{r^2}\psi \right)=0,
\end{eqnarray}
\end{subequations}
where the boundary conditions for the magnetic part must be imposed.
For given $\eta$, we can seek the minimum of eigenvalue $R_m$ over $k,A$. The result is shown in figure 11.

Some further analytical progress can be made for the narrow-gap cases.
For both of the boundary conditions the optimum of $R_m$ occurs at $k\rightarrow 0$.
For the perfectly conducting walls we substitute the small $k$ expansions
\begin{eqnarray}
b=\frac{1+k^2b_1+\cdots}{iR_mA},\qquad \psi=k\psi_0+\cdots
\end{eqnarray}
to the narrow-gap version of (\ref{AAAA}). The leading order equations are
\begin{eqnarray}
\psi_0''+1=0,\qquad
(b_1''-1)+R_m^2 A (A\psi_0''+\psi_0)=0.
\end{eqnarray}
Upon using the solution of the first equation $\psi_0=(1-x^2)/2$ to the second equation, and integrating once,
\begin{eqnarray}
R_m^2 A^2-\frac{R_m^2 A}{3}+1=0.\label{cdAAAA}
\end{eqnarray}
Here we required that $\phi_0=b_0'=0$ on the walls. 
From (\ref{cdAAAA}) it is easy to find that the minimum $R_m=6$ occurs at $A=1/6$. 

For the perfectly insulating case the appropriate expansions are
\begin{eqnarray}
b=ikb_0+\cdots,\qquad \psi=1+k\psi_0+\cdots, \qquad A=\frac{A_0}{kR_m}+\cdots.
\end{eqnarray}
The leading order equations
\begin{subequations}
\begin{eqnarray}
A_0 b_0= \psi_0'',\label{fAAAA}\\
A_0\psi_0''+R_m=b_0''
\end{eqnarray}
\end{subequations}
yields
\begin{eqnarray}
b_0=\frac{R_m}{A_0^2}\left ( \frac{\text{cosh}A_0x}{\text{cosh}A_0}-1\right )
\end{eqnarray}
which satisfies $b_0=0$ on the walls. 
Substituting this solution to (\ref{fAAAA}) and integrate once we have dispersion relation
\begin{eqnarray}
R_m= \frac{A_0^2}{A_0-\text{tanh} A_0}.
\end{eqnarray}
Here we used $\psi_0'\pm 1=0$ at $x=\pm 1$. The minimum $R_m\approx 3.774$ at $A_0\approx 1.606$ is consistent to figure 11.

\end{document}